\documentclass{aa}

\usepackage{lastpage}
\usepackage{graphicx,txfonts,textcomp}
\usepackage{colortbl}
\usepackage{booktabs}
\usepackage{lipsum}   
\usepackage{amsmath}
\usepackage{cancel}
\usepackage{adjustbox}
\usepackage{makecell}
\usepackage{multirow}
\usepackage{soul}
\usepackage{xcolor}
\definecolor{DarkBlue}{rgb}{0.0,0.3,0.5} 

\usepackage[colorlinks]{hyperref}
\hypersetup{
    colorlinks = True,
    citecolor = DarkBlue,
    linkcolor = DarkBlue,
    urlcolor = DarkBlue
}

\usepackage{natbib}

\def\apjl{{Astrophys.~J.~Lett.}}
\def\apjs{{Astrophys.~J.~Supp.}}
\def\ao{{Appl.~Opt.}}

\def\aap{{Astron.~Astrophys.}}

\definecolor{electricviolet}{RGB}{143,0,255} 

\definecolor{darkorange}{RGB}{255, 140, 0} 

\newcommand{\orcidlink}[1]{\href{https://orcid.org/#1}{\includegraphics[width=10pt]{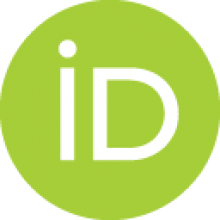}}}

\begin{document}

\title{Comparing next-generation detector configurations for high-redshift gravitational wave sources \\ with neural posterior estimation}

\titlerunning{Comparing XG detector configurations for high-$z$ GW sources using NPE}

\author{F. Santoliquido$\,$\orcidlink{0000-0003-3752-1400}\inst{1,2}\fnmsep\thanks{Corresponding author: \href{mailto:filippo.santoliquido@gssi.it}{filippo.santoliquido@gssi.it}}
    \and J. Tissino$\,$\orcidlink{0000-0003-2483-6710}\inst{1,2}
    \and U. Dupletsa$\,$\orcidlink{0000-0003-2766-247X}\inst{3}
    \and M. Branchesi$\,$\orcidlink{0000-0003-1643-0526}\inst{1,2}
    \and J. Harms$\,$\orcidlink{0000-0002-7332-9806}\inst{1,2}
    }

\institute{Gran Sasso Science Institute (GSSI), I-67100 L'Aquila, Italy
\and INFN, Laboratori Nazionali del Gran Sasso, I-67100 Assergi, Italy \and Marietta Blau Institute (MBI) - Austrian Academy of Sciences, 1010 Vienna, Austria}

\abstract 
{The coming decade will be crucial for determining the final design and configuration of a global network of next-generation (XG) gravitational-wave detectors, including the \textit{Einstein} Telescope (ET) and Cosmic Explorer (CE). In this study, and for the first time, we assessed the performance of various network configurations using neural posterior estimation (NPE) implemented in {\sc Dingo-IS}---a method based on normalizing flows and importance sampling that enables fast and accurate inference. We focused on a specific science case involving short-duration, massive and high-redshift binary black hole mergers with detector-frame chirp masses $(\mathcal{M}_\mathrm{d})$ $> 100~\mathrm{M}_\odot$. These systems encompass early-Universe stellar and primordial black holes, as well as intermediate-mass black hole binaries, for which XG observatories are expected to deliver major discoveries. Validation against standard Bayesian inference demonstrates that NPE robustly reproduces complex and disconnected posterior structures across all network configurations. For a network of two misaligned L-shaped ET detectors (2L~MisA), the posterior distributions on luminosity distance can become multimodal and degenerate with the sky position, leading to less precise distance estimates compared to the triangular ET configuration. However, the number of sky-location multimodalities is substantially lower than the eight expected with the triangular ET, resulting in improved sky and volume localization. Adding CE to the network further reduces sky-position degeneracies, and the better performance of the 2L~MisA configuration over the triangle remains evident.}


\maketitle
\nolinenumbers

\section{Introduction}

Next-generation (XG) detectors such as the \textit{Einstein} Telescope \citep[ET;][]{Punturo:2010zz,ET:2019dnz,Branchesi:2023mws} and Cosmic Explorer \citep[CE;][]{Reitze:2019iox,Evans:2023euw,Gupta:2023lga} will shape the future of gravitational-wave (GW) astronomy. 
Two of the greatest scientific breakthroughs expected with XG detectors are the observation of early-Universe binary black hole (BBH) mergers, potentially detectable up to redshifts $\lesssim 100$, and the detection of intermediate-mass black hole (IMBH) binary mergers, with source-frame total masses reaching up to $\lesssim 10^4\,\mathrm{M}_\odot$ \citep{Hall:2019xmm}.  
 High-redshift BBHs include two main types of systems: \textit{(i)}~BBHs formed from Population~III (Pop.~III) stars have been extensively and recently studied as promising high-redshift GW sources of astrophysical origin \citep{Hartwig:2016nde, Belczynski:2016ieo, kinugawa2020, liubromm2020, tanikawa2022, wang2022, Mestichelli:2024djn, Santoliquido:2024oqs}. In particular, in \cite{santoliquido2023} we found that between $\sim 20$\% and $\sim 70$\% of Pop.~III BBHs detectable with ET could merge at $z > 8$. \textit{(ii)}~Primordial black holes (PBHs), which originate from the collapse of large inhomogeneities during the radiation era \citep{Zeldovich:1967lct,Hawking:1974rv,Chapline:1975ojl,Carr:1975qj}, are also expected to form at high redshifts and span a wide mass spectrum, including the stellar-mass regime accessible to XG observatories, and a broad range of merger rates \citep{Ivanov:1994pa,Guth:2012we,Ivanov:1997ia,Blinnikov:2016bxu,DeLuca:2020qqa, Franciolini:2022tfm, Ng:2022agi,Carr:2020gox,Franciolini:2023opt}.  

In the $10^2$--$10^3$~M$_\odot$ range, IMBHs are thought to form primarily through stellar collisions \citep{2002ApJ...576..899P, 2015MNRAS.454.3150G, 2016MNRAS.459.3432M, 2021A&A...652A..54A, 2023MNRAS.526..429A}, hierarchical black hole mergers \citep{Gerosa:2017kvu, Fishbach:2017dwv, Gerosa:2021mno, 2002MNRAS.330..232C, 2004ApJ...616..221G, 2019MNRAS.486.5008A, 2021A&A...652A..54A, Kritos:2022ggc}, close interactions between stars and black holes \citep{2017MNRAS.467.4180S, 2023MNRAS.521.2930R, 2023MNRAS.526..429A}, or a combination of these mechanisms \citep{2015MNRAS.454.3150G, 2023MNRAS.526..429A}. Accurate parameter estimation is essential for uncovering the origins of early-Universe BBHs and for determining the relative importance of the various formation pathways of IMBHs. However, the high expected detection rate of XG observatories, about
$10^5$ events per year 
\citep{Baibhav:2019gxm}, poses unprecedented challenges for parameter estimation and data analysis \citep[see also chapter 10 of][]{Abac:2025saz}.

To handle the vast amount of data expected, it is crucial to accelerate parameter estimation for individual sources, as current stochastic sampling methods, often requiring several hours per event \citep{Smith:2019ucc}, are inadequate for the era of XG detectors. The acceleration of parameter estimation can be pursued through three complementary strategies: simplifying the likelihood function \citep{2010arXiv1007.4820C,Veitch:2014wba,Zackay:2018qdy,2021PhRvD.104l3030L,Narola:2023men,Vinciguerra:2017ngf,Aubin:2020goo, 2021PhRvD.104d4062M,Allene:2025saz,2015PhRvL.114g1104C,Smith:2021bqc,2023PhRvD.107h4037T,Morras:2023pug,Smith:2021bqc,Narola:2024qdh,2024arXiv241202651H}, improving sampling efficiency \citep{Williams:2021qyt,2023MLS&T...4c5011W,nessai,2025MNRAS.541..200P,2023JOSS....8.5021W,2023ApJ...958..129W,2024PhRvD.110h3033W,Perret:2025wox,Negri:2025cyc,Prathaban:2025qgg}, and employing 
methods based on deep neural networks \citep{Cuoco:2020ogp}.

In recent years, likelihood-free inference (or simulation-based inference)---particularly neural posterior estimation \citep[NPE;][]{Green:2020hst,Green:2020dnx,Dax:2021myb,Chatterjee:2022ggk,2023mla..confE..34W,2023PhRvL.130q1402L,Dax:2022pxd,DeSanti:2024oap,Lanchares:2025kpb,Marx:2025ioo,Srinivasan:2025etu,Kofler:2025dux,Chan:2025pdf}---has enabled much faster and accurate parameter estimation, producing posterior samples within minutes. \cite{Dax:2021tsq} implemented NPE in {\sc Dingo} using conditional normalizing flows \citep{Papamakarios:2019fms,Kobyzev_2021}, which are invertible neural networks that transform a base distribution, typically a standard normal, into a more complex distribution, which serves as an approximation of the posterior. 
\cite{Santoliquido:2025lot} trained {\sc Dingo} on the triangular configuration of ET and showed that NPE performs effectively in the XG-detector regime, delivering accurate parameter estimation with excellent computational efficiency. 
 
The science goals of XG observatories critically depend on design choices that are effectively irreversible, including the detectors' geometry, geographic location, and orientation \citep[e.g.,][]{Branchesi:2023mws,Gupta:2023lga}. 
Making the right decisions at the design stage is therefore crucial to maximizing the scientific potential of XG detectors. 

This work presents the first quantitative assessment of how different XG detector configurations affect the parameter estimation performance for massive BBH mergers ($\mathcal{M}_\mathrm{d} > 100~\mathrm{M}_\odot$), based on full Bayesian analyses. Leveraging NPE enables the accurate evaluation of key metrics, including localization volumes and the presence of multimodal sky posteriors.

This manuscript is structured as follows. Section \ref{sec:detectors} outlines the examined detector configurations, while Sect. \ref{sec:dingo} introduces {\sc Dingo-IS}, which combines NPE with importance sampling. Section \ref{sec:metrics} defines the performance metrics used throughout the work. Section \ref{sec:high_eff} shows that {\sc Dingo-IS} correctly recovers a large fraction of injections, between 85\% and 96\% depending on the detector configuration. Section \ref{sec:single} presents a single-injection study showing that a standard inference method and {\sc Dingo-IS} yield statistically indistinguishable posteriors, with the latter requiring far less computational time. Section \ref{sec:pe_perfom} compares parameter-estimation performance across detector configurations, highlighting that two misaligned L-shaped ET detectors outperform the triangular ET for sky localization and volume. Finally, Sect. \ref{sec:conclusions} summarizes our conclusions.

\section{Methods}

\subsection{Detector and network configurations}
\label{sec:detectors}

We considered seven different XG detector configurations:
\begin{itemize}
    \item {$\Delta$}. One single triangular ET with 10 km arms located in Sardinia \citep{Branchesi:2023mws}.
    \item 2L~A. Two L-shaped ET detectors, each with 15 km arms, positioned in Sardinia and in the Meuse-Rhine Euro (EMR) region, with parallel arms \citep{Branchesi:2023mws}.
    \item 2L~MisA. Same as 2L~A but with the detector arms misaligned by 45° \citep{Branchesi:2023mws}.
    \item 2L~MisA~+~LHI. Same as the 2L~MisA configuration but including three LIGO observatories---Livingston, Hanford, and India \citep{Unnikrishnan:2013qwa}---operating at the mid-2030s A\# LIGO sensitivity.
    \item 1L~+~CE. An L-shaped ET detector with 15 km arms in Sardinia combined with a single CE detector with 40 km arms located near Hanford \citep{Reitze:2019iox}. This configuration, introduced in \cite{Maggiore:2024cwf} but not adopted as an official design by the ET Collaboration, is included here solely for comparison purposes.
    \item $\Delta$~+~CE. A single triangular ET detector located in Sardinia with 10 km arms observing  with CE, as defined in 1L~+~CE.
    \item 2L~MisA~+~CE. Same as 1L~+~CE but with a second L-shaped ET detector located in the EMR region, rotated by 45° relative to the first ET.
\end{itemize}

During the development of this work, the Lusatia region in Germany, near Kamenz (51.275°, 14.100°), was announced as the third official candidate site for ET. Its chord distance from the Sardinian site is approximately 1248 km, only about 82 km longer than the distance between the EMR region and Sardinia. Given this minimal difference, we consider the two sites effectively equivalent for hosting the second L-shaped interferometer for the specific science case considered here.

Additional details, including coordinates, chord distances between interferometers, and further information on the amplitude spectral densities (ASDs), are provided in Appendix~\ref{app:detectors}.

\subsection{{\sc{Dingo-IS}}}
\label{sec:dingo}

We conducted our analysis using NPE implemented in {\sc Dingo}, where a conditional probabilistic neural network ($q_\phi(\theta | d)$) with tunable parameters ($\phi$) approximates the true posterior ($p(\theta | d)$). {\sc Dingo} is trained on simulated datasets $(\theta,d)$, where the parameters ($\theta$) are sampled from the priors and the corresponding data ($d$) consist of a GW signal embedded in stationary Gaussian noise, $d=h(\theta)+n$. The prior distribution is the same as that adopted in \cite{Santoliquido:2025lot} and is listed in Appendix~\ref{app:prior}, along with the choices of waveform parameters and approximants.

Compared to the network architecture in \cite{Santoliquido:2025lot}, where all {\sc Dingo} hyperparameters were kept at their default values, we enlarged the embedding network that compresses the input strain data into a set of features. Specifically, the output dimensionality was increased from 128 to 256 features, and several additional hidden layers were added, resulting in a total of $4 \times 10^8$ learnable parameters. Appendix~\ref{app:learning} shows the learning curves for each considered detector configuration, along with the corresponding probability--probability plots.

After training, {\sc Dingo} can rapidly generate approximate posterior samples. These samples, however, can deviate from the true posterior. A practical way to assess and refine them is via importance sampling \citep{https://doi.org/10.1002/wics.56, Payne:2019wmy, Elvira_2021, Ashton:2025xba}. Combining this technique with {\sc Dingo} leads to {\sc Dingo-IS} \citep{Dax:2022pxd}, which assigns weights to a set of $n$ samples of $\theta_i$ drawn from the proposal distribution $q_\phi(\theta|d)$:
\begin{equation}
\label{eq:weight}
    w_i = \frac{\mathcal{L}(d|\theta_i)\pi(\theta_i)}{q_\phi(\theta_i|d)},
\end{equation}where $\mathcal{L}(d|\theta)$ is the likelihood function defined in Appendix~\ref{app:bilby}. 
Ideally, all weights ($w_i$) would be equal; in practice, the sample efficiency quantifies how well $q_\phi(\theta|d)$ approximates the target distribution:
\begin{equation}
\label{eq:eff}
    \epsilon = \frac{n_{\mathrm{eff}}}{n} \in \left[\frac{1}{n},\,1\right],
\end{equation}with the effective number of samples given by $n_\mathrm{eff} = (\sum_i w_i)^2 / \sum_i w_i^2$ \citep{Kong}. The lower bound $\epsilon=1/n$ corresponds to the extreme case where one weight dominates and all others are negligible, which is the minimum possible as the weights are normalized to have a mean equal to 1. We compared the results obtained with \textsc{Dingo-IS} to those derived using standard inference methods implemented in \textsc{Bilby} (see Appendix~\ref{app:bilby} for additional details).

\subsection{Metrics to assess detector performance}
\label{sec:metrics}

We defined the metrics to evaluate the performance in estimating parameters of the various configurations of XG detectors. The information gain ($I$) is defined as the Kullback--Leibler (KL) divergence \citep[][]{10.1214/aoms/1177729694} between the prior and the posterior \citep{2022RNAAS...6...89B}.  
In particular, we computed the KL divergence of the posterior from the prior:
\begin{equation}
\label{eq:info_gain} 
 I = \int p(\theta|d) \log_2 \frac{p(\theta|d)}{\pi(\theta)} \mathrm{d}\theta,  
\end{equation}which quantifies the extent to which the prior volume has been constrained by the data, i.e., by a factor of $2^I$ \citep{Skilling:2004pqw}. As opposed to single- or few-parameter metrics, the information gain fully characterizes the size of the 11-dimensional posterior distribution, accounting for correlations across any parameters. Equation~\ref{eq:info_gain} was evaluated using a Monte Carlo integral. Additional details on this are provided in Appendix~\ref{app:info}.

Because {\sc Dingo-IS} provides full parameter estimates for each source, we can define precise metrics to assess how different detector networks perform in source localization. For each event, we computed the area (in deg$^2$) enclosed by the smallest 90\% confidence region. The sky maps were sampled using equal-area HEALPix (Hierarchical Equal Area isoLatitude Pixelization) pixels \citep[][]{Gorski:1999rt,Gorski:2004by}, where each of the $N$ pixels carries a posterior probability ($p_i$) that the source lies within that pixel. 

We employed the {\texttt{ligo-skymap}} library \citep{Singer:2015ema, Singer:2016eax, Singer:2016erz}, which also identifies the number of disconnected sky modes (or probability islands) within a given area using a flood-fill algorithm \citep{foley1996computer}, which finds all pixels connected to a starting one. Additionally, by including the posterior on luminosity distance, we computed the 90\% highest probability density comoving volume \citep[$\Delta V^{c}_{90\%}$; see Sect.~3 of][]{Singer:2016eax}.

The other reported interval widths, for example $\Delta x/x^{\mathrm{inj}}$, indicate the relative variation of a parameter from its injected value. Here, $\Delta x$ corresponds to the square root of the associated covariance matrix element and represents the $1\sigma$ uncertainty of the parameter estimate.

\section{Results}
\label{sec:results}

\subsection{High signal-to-noise ratio versus high sample efficiency}
\label{sec:high_eff}

\begin{figure}
    \centering
    \includegraphics[width=\linewidth]{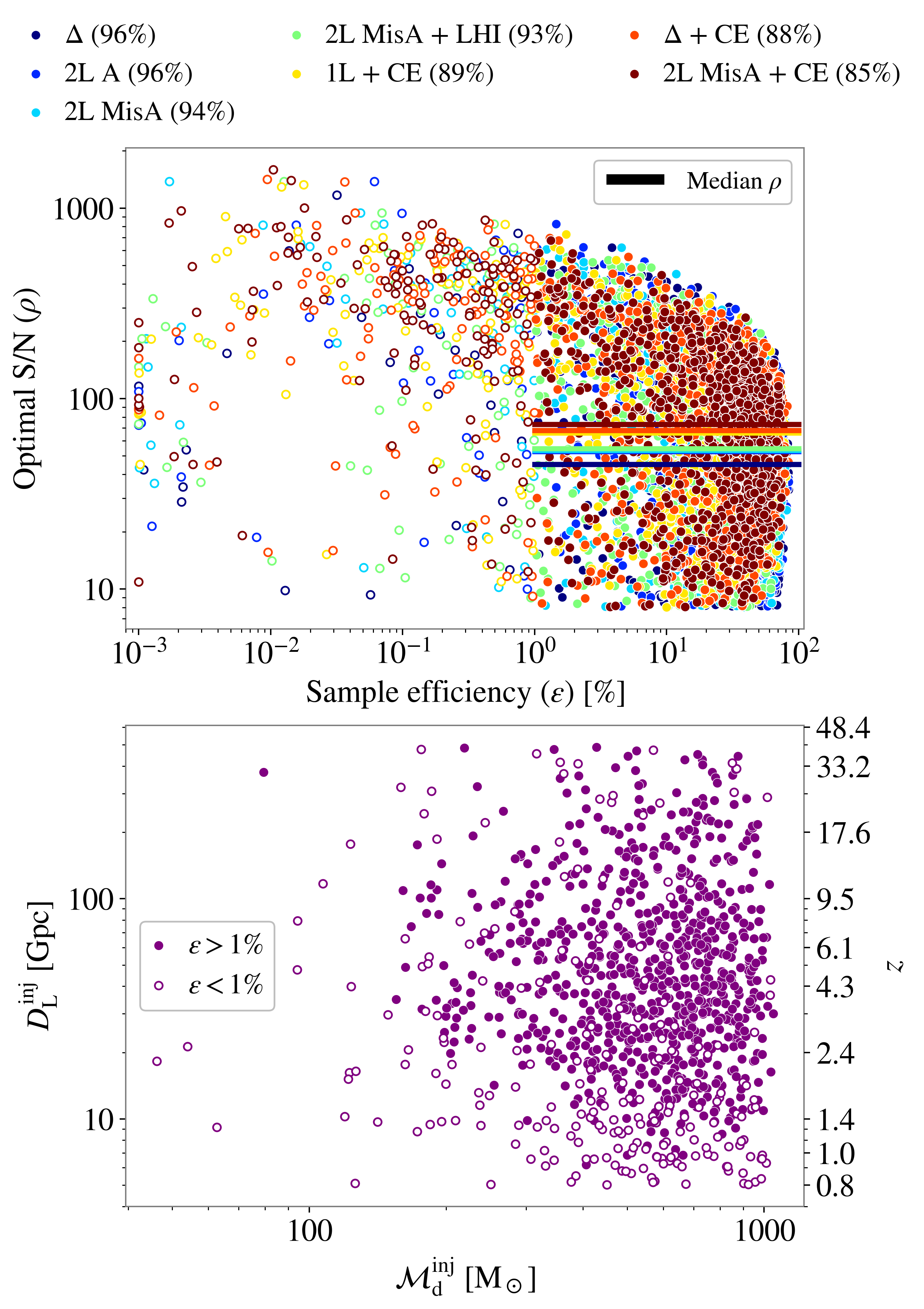}
    \caption{\textit{Top panel}: Injections as a function of sample efficiency ($\epsilon$, $x$-axis) and  optimal S/N ($\rho$, $y$-axis) for each  detector configuration (color-coded). Filled markers indicate injections with sample efficiencies $> 1\%$. Percentages in parentheses denote the fraction of sources with a sample efficiency above $1\%$, while the horizontal colored lines indicate the median $\rho$ for sources exceeding this threshold. \textit{Bottom panel}: Injections as a function of detector-frame chirp mass ($x$-axis) and luminosity distance (left $y$-axis), with corresponding redshifts (right $y$-axis). Colored markers indicate events with a sample efficiency greater than $1\%$ in all considered detector configurations. See Sect.~\ref{sec:high_eff} for details.}
    \label{fig:SNR_vs_eff}
\end{figure}

We generated $1000$ sets of BBH merger parameters sampled from the prior and used the same samples for all configurations. From the {\sc Dingo} proposal distribution $q_\phi(\theta|d)$, we drew $10^5$ posterior samples and computed the corresponding importance weights (see Sect.~\ref{sec:dingo}). For each injected signal, we then evaluated the sample efficiency. To ensure unbiased estimates of the inferred parameters, we required at least $10^3$ effective samples per source, corresponding to a sample efficiency $> 1$ \%. 

The top panel of Fig.~\ref{fig:SNR_vs_eff} shows that a large fraction of injected signals achieves a sample efficiency greater than $1\%$, ranging from $96\%$ for $\Delta$ and 2L~A configurations to a minimum of $85\%$ for 2L~MisA~+~CE. These values remain consistently high, with even better performance for the $\Delta$ configuration than previously reported in \cite{Santoliquido:2025lot}. This improvement arises from the increased dimensionality of the embedding network (see Sect.~\ref{sec:dingo}), emphasizing the crucial role of this component within the NPE framework.

The top panel of Fig.~\ref{fig:SNR_vs_eff} also shows that $50\%$ of the events have an optimal signal-to-noise ratio (S/N; see Appendix~\ref{app:bilby}) exceeding $40$ for the $\Delta$ configuration and up to $70$ for 2L~MisA~+~CE, while still achieving sample efficiencies above $1\%$. This demonstrates that the sample efficiency remains high even for high-S/N events, indicating that NPE may represent a promising approach to addressing the parameter estimation challenges posed by XG detectors.

To ensure an unbiased evaluation of parameter estimation performance, we included only injections with sample efficiencies exceeding $1\%$ across all seven detector configurations. We further restricted the dataset to sources with optimal S/N $\rho > 8$. The bottom panel of Fig.~\ref{fig:SNR_vs_eff} illustrates the parameter space of the retained sources, which account for 78\% of the total, showing that events with $\mathcal{M}_d \lesssim 150\,\mathrm{M}_\odot$ and $D_{\mathrm{L}} \lesssim 6$~Gpc are almost entirely excluded. These sources have low sample efficiencies because their high S/N makes them more challenging for NPE to model, as their posteriors are much narrower than the prior. This behavior is further evident in the top panel of Fig.~\ref{fig:SNR_vs_eff}, where sample efficiency drops sharply for sources with optimal S/Ns of around $\sim 1000$.

We are currently exploring several strategies to improve sample efficiency in regions of parameter space where performance remains below the threshold, including training {\sc Dingo-IS} with neural population priors \citep[e.g.,][]{Wouters:2025csq}. The results of these efforts will be presented in future work.

\subsection{Multimodal posterior of a single event}
\label{sec:single}

\begin{figure}
    \centering
    \includegraphics[width=\linewidth]{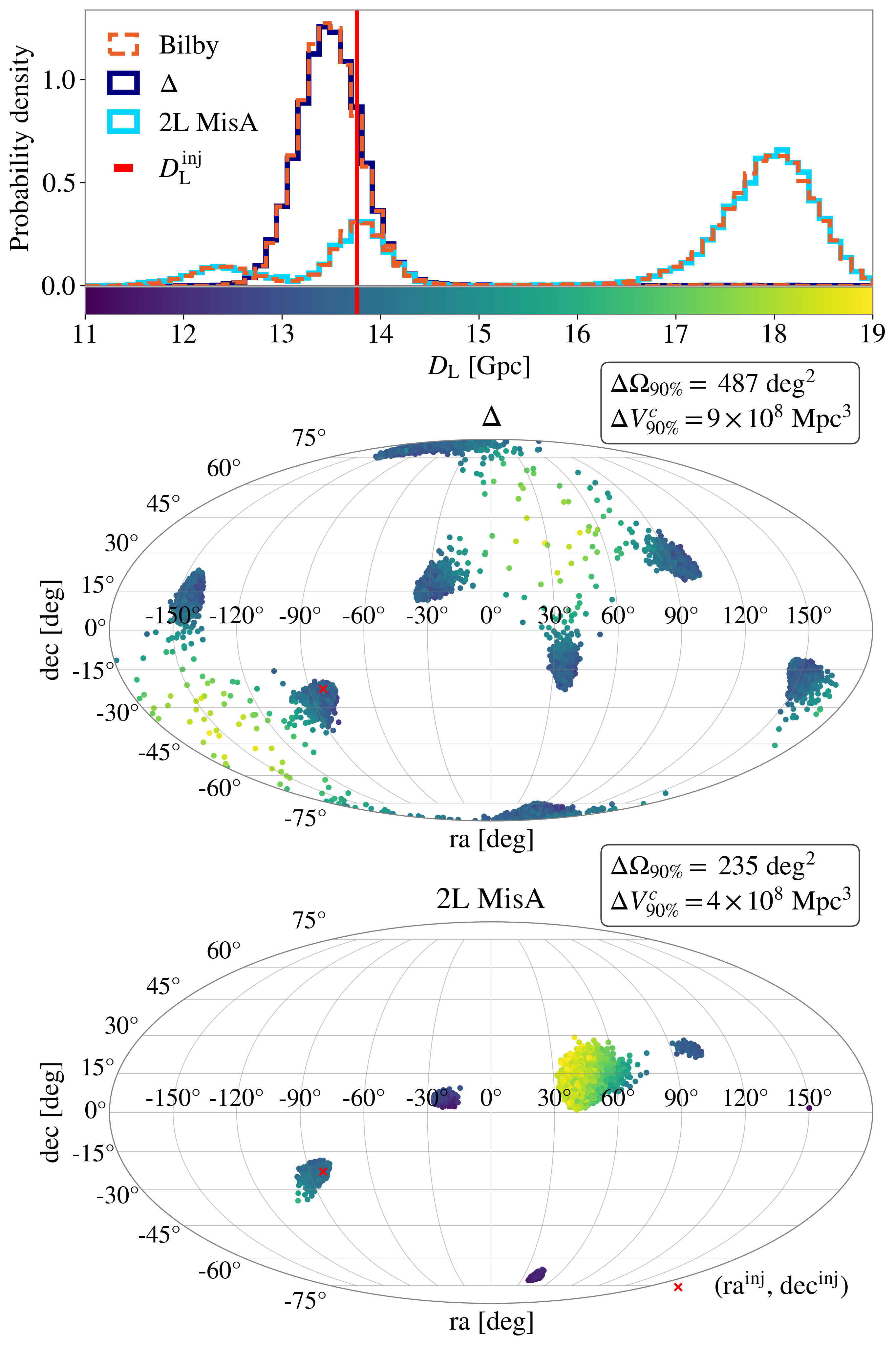}
    \caption{\textit{Top panel:} Marginalized one-dimensional posterior distributions for the luminosity distance, recovered with \textsc{Dingo-IS,} for an event observed with the $\Delta$ (dark blue) and 2L~MisA (light blue) configurations. The results are compared with \textsc{Bilby} (orange). The vertical line marks the injected luminosity distance. The remaining parameters are shown in Appendix~\ref{app:bilby}. \textit{Middle panel:} Posterior samples obtained with \textsc{Dingo-IS} in the $\Delta$ configuration for right ascension (ra) and declination (dec), color-coded by luminosity distance. The red cross marks the injected sky position. We also report the sky-localization area ($\Delta \Omega_{90\%}$) and comoving-volume localization ($\Delta V^c_{90\%}$) errors. \textit{Bottom panel:} Same as the middle panel but for the 2L~MisA configuration. See Sect.~\ref{sec:single} for details.}
    \label{fig:dingo_bilby_2L_short}
\end{figure}

The top panel of Fig.~\ref{fig:dingo_bilby_2L_short} shows that \textsc{Dingo-IS} yields posterior distributions that are statistically indistinguishable from those obtained using \textsc{Bilby} (see Appendix~\ref{app:bilby} for a formal validation). Crucially, this accuracy is achieved at a dramatically lower computational cost: \textsc{Dingo-IS} requires only a few minutes per source, whereas \textsc{Bilby} needs more than 22 hours (50 hours) to converge for this event with the $\Delta$ (2L~MisA) configuration.

Figure~\ref{fig:dingo_bilby_2L_short} also shows that \textsc{Dingo-IS} accurately captures the complex structure of multimodal posteriors. These multimodalities include both the eight-fold sky degeneracy arising from the triangular geometry of ET, extensively discussed in \cite{Santoliquido:2025lot}, and modes introduced by the 2L~MisA configuration, which affect both sky localization and luminosity distance. 

The bottom panel of Fig.~\ref{fig:dingo_bilby_2L_short} shows that for the 2L~MisA configuration, the sky location is correlated with luminosity distance: distinct islands of probability on the sky correspond to different modes in $D_{\mathrm{L}}$. For this source, we identify three such modes. The occurrence rate of multimodal posteriors in luminosity distance is discussed in Appendix~\ref{app:multidl} and shown in Fig.~\ref{fig:diptest}. The 2L~MisA (2L~A) configurations produce a multimodal $D_{\mathrm{L}}$ posterior more frequently than the $\Delta$ configuration, with rates of about 20\% compared to $\sim$2\%. In Appendix~\ref{app:waveforms} we show that the multimodalities in luminosity distance and sky localization observed for the 2L configurations are robust features, persisting across different waveform approximants.

These multimodalities are expected when observing short-duration, high-mass BBHs with both the 2L~A and 2L~MisA configurations of ET. They arise from the combined effect of the short baseline between the two interferometers ($\sim$1000 km) and the low merger frequencies of massive BBHs (typically $\lesssim 50$ Hz). We provide an illustrative example in Appendix~\ref{sec:time}, showing that for a signal with higher merger frequencies, the multimodalities in both sky position and distance disappear.

Despite the multimodality in $D_{\mathrm{L}}$, the number of sky-position modes is significantly reduced compared to the triangular configuration, resulting in improved sky-area and comoving-volume localization for the 2L~MisA configuration. Additional details, including the same source as observed with the 2L~MisA~+~CE configuration, are provided in Appendix~\ref{app:bilby}.

\subsection{Parameter estimation performance}
\label{sec:pe_perfom}

\begin{figure*}
    \centering
    \includegraphics[width=0.9\linewidth]{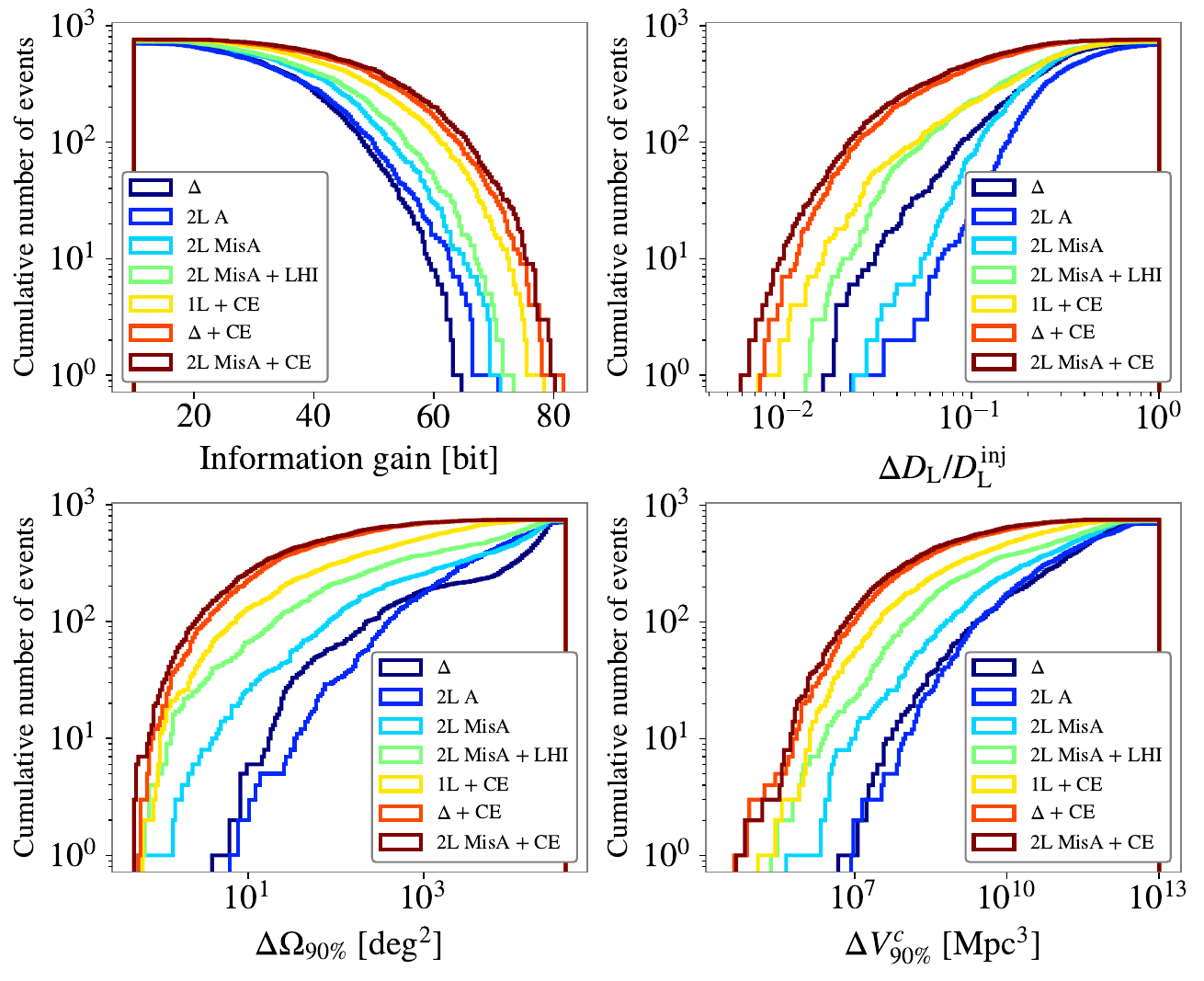}
    \caption{Cumulative distributions of events as a function of the information gain (\textit{top-left panel}; see also Sect.~\ref{sec:metrics}), the relative variation in luminosity distance (\textit{top-right panel}; $\Delta D_{\mathrm{L}}/D^{\mathrm{inj}}_{\mathrm{L}}$), sky (\textit{bottom-left panel}; $\Delta \Omega_{90\%}$; see also Table~\ref{tab:sky}), and comoving volume localization (\textit{bottom-right panel}; $V^{c}_{90\%}$) for all considered XG detector configurations (color-coded). See Sect.~\ref{sec:pe_perfom} for details.}
    \label{fig:bigpanel}
\end{figure*}

The top-left panel of Fig.~\ref{fig:bigpanel} shows the parameter estimation performance of the considered detector configurations, measured by the information gain, revealing a clear ranking. Among the ET configurations ($\Delta$, 2L~A, and 2L~MisA), when no other XG detectors are observing simultaneously, the 2L~MisA setup performs best. Moreover, adding three LIGO observatories operating at A\# sensitivity to the 2L~MisA configuration further increases the information gain, underscoring the capability of this global network to estimate the parameters of the specific sources examined in this study---namely, both massive and high-redshift BBHs. The information gain further increases when 1L ET observes jointly with CE, as the longer baseline enhances both sky and volume localization. However, the separation between the 2L~MisA detectors also plays a role: when observing together with CE, the 2L~MisA~+~CE configuration outperforms $\Delta$~+~CE.

The top-right panel of Fig.~\ref{fig:bigpanel} shows that the luminosity distance is best measured with the $\Delta$ configuration, outperforming both the 2L~A and 2L~MisA configurations.  
Compared to \citet[see their Fig.~5]{Branchesi:2023mws}, we find a reversed hierarchy in detector performance. We stress that our analysis is restricted to both massive and high-redshift BBH mergers. For this class of source, the 2L configurations give rise to multimodal posteriors in $D_\mathrm{L}$ (see Fig.~\ref{fig:dingo_bilby_2L_short} and Appendix~\ref{app:multidl}), an effect that is not captured with Fisher-matrix-based analyses.

Nevertheless, the 2L~MisA configuration leads to better sky and volume localization, allowing it to outperform the $\Delta$ and 2L~A setup in these metrics (bottom panels of Fig.~\ref{fig:bigpanel}). Improving the accuracy of the localization volume critically enhances the prospects for dark siren cosmology, which relies on the statistical association between the GW source and galaxies contained within the reconstructed localization volume \citep{Schutz:1986gp,DelPozzo:2011vcw,Chen:2017rfc,Libanore:2020fim,Gair:2022zsa,Bosi:2023amu,Borghi:2023opd}.

\begin{table}
\caption{Percentage of injections (color-coded) with estimated sky localization errors within the specified areas (columns) for different detector configurations (rows).}
\label{tab:sky}
\centering
\begin{tabular}{lccc}
\hline\hline
\multicolumn{1}{r}{$\Delta \Omega_{90\%}$}  & $<10$ deg$^2$ & $<100$ deg$^2$ & $<1000$ deg$^2$ \\
\hline
$\Delta$ & \cellcolor{blue!0}0.7\% & \cellcolor{blue!5}8.5\% & \cellcolor{blue!15}26.2\% \\
2L A & \cellcolor{blue!0}0.3\% & \cellcolor{blue!2}4.4\% & \cellcolor{blue!14}24.3\% \\
2L MisA & \cellcolor{blue!2}3.6\% & \cellcolor{blue!9}15.0\% & \cellcolor{blue!20}34.8\% \\
2L MisA + LHI & \cellcolor{blue!5}9.5\% & \cellcolor{blue!17}29.6\% & \cellcolor{blue!30}50.9\% \\
1L + CE & \cellcolor{blue!10}17.3\% & \cellcolor{blue!25}42.2\% & \cellcolor{blue!41}69.7\% \\
$\Delta$ + CE & \cellcolor{blue!18}31.0\% & \cellcolor{blue!40}68.0\% & \cellcolor{blue!55}93.2\% \\
2L MisA + CE & \cellcolor{blue!22}36.7\% & \cellcolor{blue!42}71.5\% & \cellcolor{blue!56}94.2\% \\
\hline
\end{tabular}
\tablefoot{See Sect.~\ref{sec:pe_perfom} for details.}
\end{table}

Table~\ref{tab:sky} summarizes the sky localization performance for all detector configurations, showing a clear improvement from the 2L~A configuration, which performs worst, to 2L~MisA~+~CE, which achieves the best results. In the latter case, more than 70\% of sources are localized within 100~deg$^2$.

\begin{figure}
    \centering
    \includegraphics[width=\linewidth]{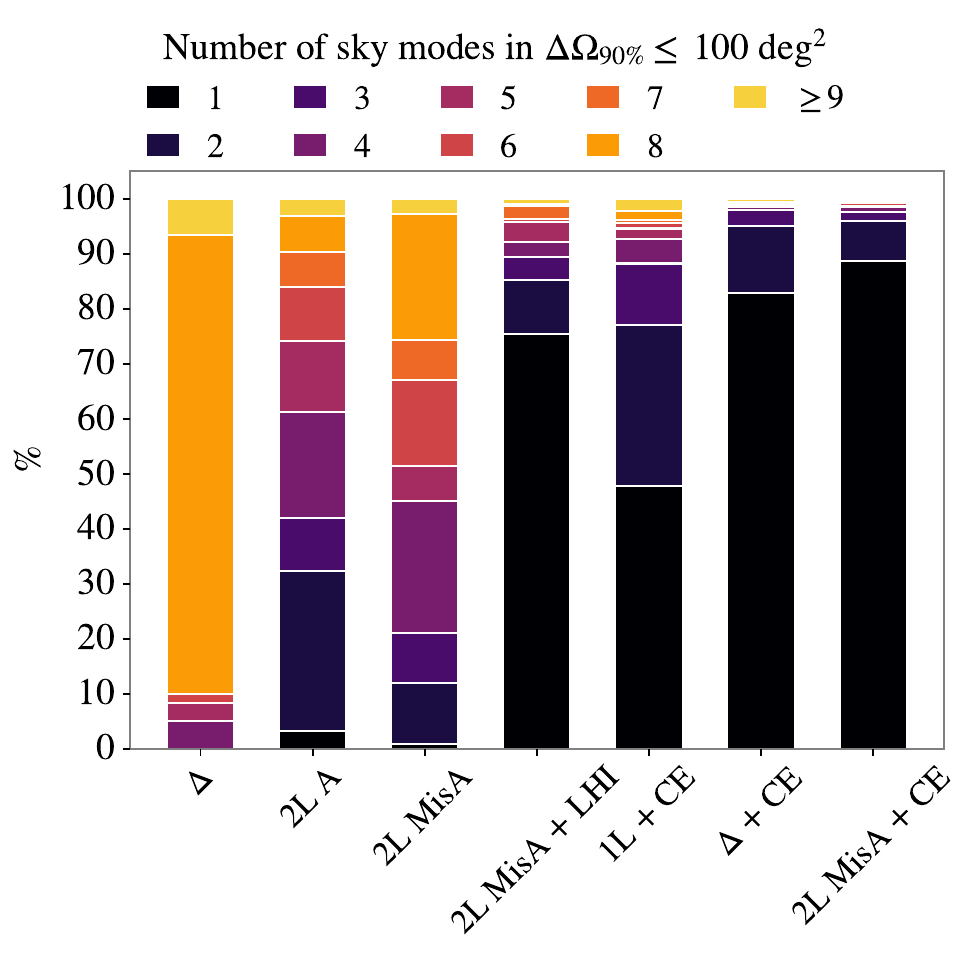}
    \caption{Percentage of sky maps ($y$-axis) with $\Delta \Omega_{90\%} \leq 100$ deg$^2$ exhibiting one to eight or more disconnected modes (color-coded), for all detector configurations ($x$-axis). See Sect.~\ref{sec:pe_perfom} for details.}
    \label{fig:modes}
\end{figure}

Figure~\ref{fig:modes} shows that more than 80\% of events exhibit eight sky modes in the $\Delta$ configuration. This fraction is substantially reduced for the two L-shaped detector configurations: in the 2L~MisA setup, the occurrence of eight sky modes drops to about 20\% of the events. The 2L~MisA configuration exhibits an eight-sky-mode pattern for events located close to the local horizon defined at the midpoint between the two L-shaped interferometers. For such events, the time-of-arrival difference is negligible, effectively making the two interferometers colocated \citep{Marsat:2020rtl}. 

The 2L~A setup localizes nearly 30\% of the sources in two distinct sky modes, outperforming the 2L~MisA configuration, for which this fraction remains below 10\%. However, the average sky-localization area per mode is significantly larger in 2L~A compared to 2L~MisA. None of the ET configurations are able to localize events in a single sky mode unless ET operates as part of a global network that includes either the three LIGO interferometers at A\# sensitivity (including LIGO--India) or CE.

\begin{figure*}
    \centering
    \includegraphics[width=0.7\linewidth]{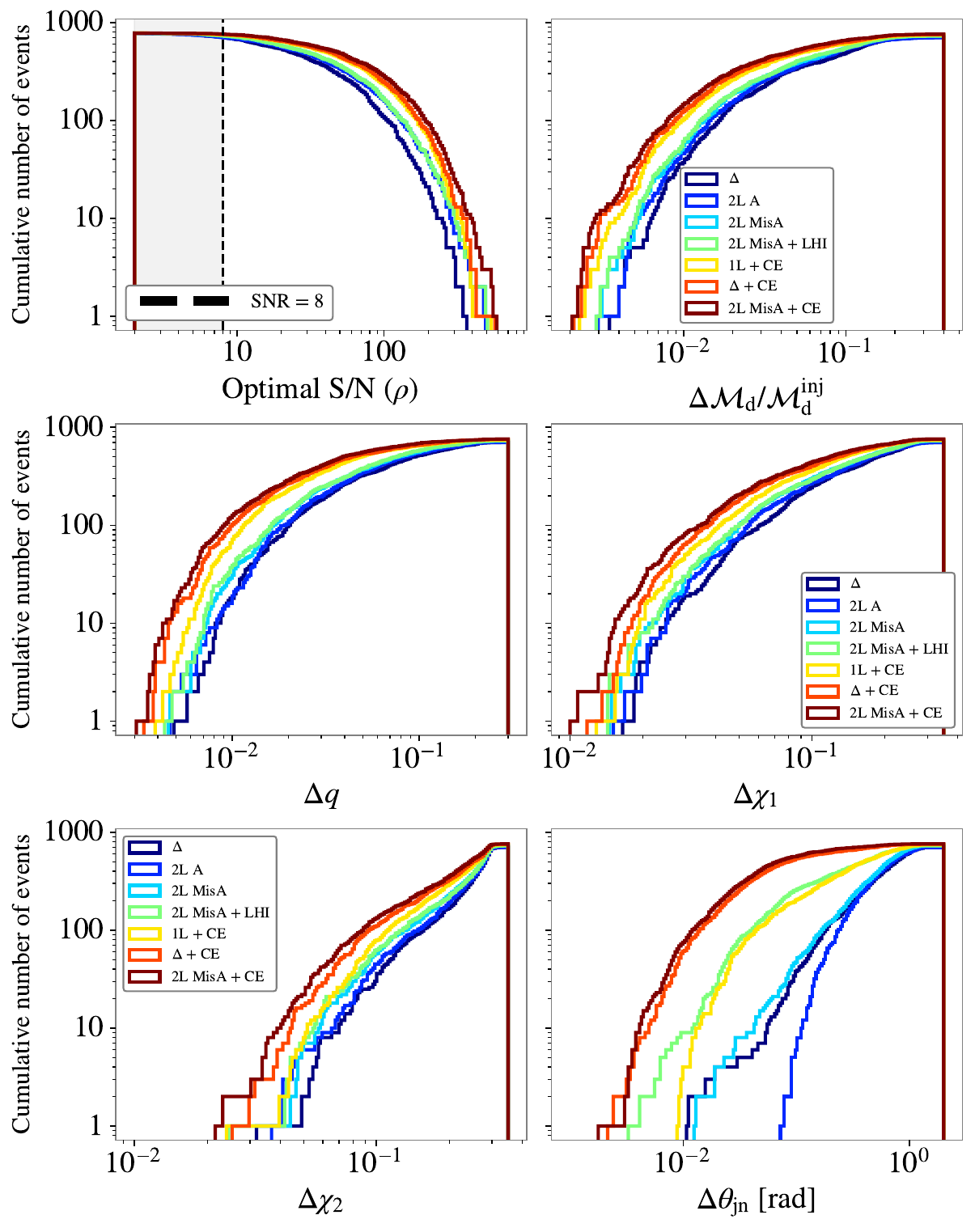}
    \caption{Cumulative distributions of events as a function of the optimal S/N ($\rho$; Eq.~\ref{eq:snr}), the relative variations in detector-frame chirp mass ($\Delta \mathcal{M}_{\mathrm{d}}/\mathcal{M}^{\mathrm{inj}}_{\mathrm{d}}$), mass ratio ($q$), first ($\Delta \chi_1$) and second ($\Delta \chi_2$) aligned spins, and inclination angle ($\Delta \theta_{\mathrm{jn}}$) for all considered XG detector configurations. See Sect.~\ref{sec:pe_perfom} for details.}
    \label{fig:bigpanelapp}
\end{figure*}

For completeness, Fig.~\ref{fig:bigpanelapp} presents the performance of the considered detector configurations in estimating the remaining parameters, such as optimal S/N, detector-frame chirp mass, mass ratio, the aligned spin components, and inclination angle. The overall trend is confirmed, as the $\Delta$ and 2L~A configurations provide the poorest performance, while the 2L~MisA~+~CE setup delivers the most accurate estimates.

The performance of ET configurations in estimating parameters of compact binary coalescences arises from the interplay between polarization measurements and a baseline that enables triangulation \citep{Fairhurst:2009tc}. Breaking the degeneracy between $D_\mathrm{L}$ and $\theta_{\mathrm{jn}}$ requires measuring both polarizations \citep{Usman:2018imj}, and the ability to do so is quantified by the alignment factor \citep{Klimenko:2006rh,deSouza:2023gjv,Mascioli:2025cnw}. A network composed of two L-shaped interferometers that are perfectly aligned provides no information on the cross-polarization. In contrast, the $\Delta$ configuration exhibits the highest alignment factor, yielding the most accurate polarization measurements. However, the sky maps produced by this configuration feature eight distinct sky modes owing to the colocated nature of the detectors \citep{Singh:2020wsy,Singh:2021bwn,Santoliquido:2025lot}. As a result, the 2L~MisA configuration offers a tradeoff, enabling polarization measurements while reducing the number of sky modes (see Fig.~\ref{fig:modes}).

\section{Conclusions}
\label{sec:conclusions}

Given the unique sensitivity that ET and other XG GW detectors will provide for early-Universe stellar and PBHs, as well as IMBH binaries, we focused our analysis on BBH mergers with detector-frame chirp masses ($\mathcal{M}_\mathrm{d}$)  $> 100~\mathrm{M}_\odot$. Although this study addresses a more restricted science case than the full range of sources observable with XG detectors \citep{Branchesi:2023mws,Abac:2025saz}, it targets a regime with strong potential for breakthrough discoveries.

We assessed the performance of seven configurations of XG detectors using NPE enhanced with importance sampling, as implemented in {\sc Dingo-IS} \citep{Dax:2022pxd}. We conduced a large injection campaign by sampling 1000 BBH merger parameters from the prior. As shown in Fig.~\ref{fig:SNR_vs_eff}, {\sc Dingo-IS} achieves excellent performance, recovering 85--96\% of signals with a sample efficiency above 1\%, depending on the detector configuration, for median S/Ns between 40 and 70. This demonstrates that its efficiency remains high even for high-S/N events. 

In Sect.~\ref{sec:single} we present a single-injection study showing that {\sc Dingo-IS} accurately reproduces complex, disconnected posteriors consistent with standard stochastic-sampling methods but at a fraction of the computational cost---a few minutes instead of several hours. Our analysis shows that a network of two misaligned L-shaped ET detectors (2L~MisA) can produce multimodal luminosity--distance posteriors (see the top panel of Fig.~\ref{fig:dingo_bilby_2L_short}), and its distance estimates are generally less precise than those obtained with the triangular ET design ($\Delta$). However, Fig.~\ref{fig:bigpanel} demonstrates that the 2L~MisA  outperforms the $\Delta$ configuration in terms of information gain, sky localization, and volume reconstruction. The $\Delta$ configuration's larger number of sky modes leads to poorer sky-localization performance, which in turn dominates its volume uncertainty.

For both high-mass and high-redshift BBH events, the fraction of sources recovered within a single sky mode remains low for all ET configurations ($\Delta$, 2L~A, and 2L~MisA). It surpasses 50\% when CE is added to the global XG network and exceeds 70\% when, instead, three LIGO detectors operate at A\# sensitivity, as shown in Fig. \ref{fig:modes}.

\section*{Data availability}

The data used in this study are publicly available on Zenodo. The dataset corresponding to the $\Delta$ configuration can be found at \citet[\href{https://zenodo.org/records/18911700}{https://zenodo.org/records/18911700}]{santoliquido_2026_18911700}; the datasets for the 2L~A, 2L~MisA, and $\Delta$~+~CE configurations are available at \citet[\href{https://zenodo.org/records/18911495}{https://zenodo.org/records/18911495}]{santoliquido_2026_18911495}; and the datasets for 2L~MisA~+~LHI, 1L~+~CE, and 2L~MisA~+~CE are available at \citet[\href{https://zenodo.org/records/18911380}{https://zenodo.org/records/18911380}]{santoliquido_2026_18911380}. The latest public release of {\sc Dingo} is available at \href{https://github.com/dingo-gw/dingo}{https://github.com/dingo-gw/dingo}; this work is based on commit \href{https://github.com/filippo-santoliquido/dingo-ET/commit/df6e586bd7efe7b93f81d02ee66c0dfd3c60c957}{{\sc df6e586}} in the fork {\sc Dingo-ET} (\href{https://github.com/filippo-santoliquido/dingo-ET}{https://github.com/filippo-santoliquido/dingo-ET}). Additional data and code are available from the corresponding authors upon reasonable request.

\begin{acknowledgements}

We thank the anonymous referee for their thorough and careful review of our manuscript. 
We thank Stephen R. Green, Francesco Iacovelli, Luca Reali, Alexandre Toubiana, Manuel Arca Sedda, Michele Maggiore, Archisman Gosh, Elisa Maggio and Emanuele Berti   
for useful discussions. F.S. has been funded by the European Union -- NextGenerationEU under the Italian Ministry of University and Research (MUR) "Decreto per l’assunzione di ricercatori internazionali post-dottorato PNRR" - Missione 4 "Istruzione e Ricerca" Componente 2 "Dalla Ricerca all'Impresa" del PNRR - Investimento 1.2 “Finanziamento di progetti presentati da giovani ricercatori” - CUP D13C25000700001. M.B. and J.H. acknowledge support from the Astrophysics Center for Multi-messenger Studies in Europe (ACME), funded under the European Union’s Horizon Europe Research and Innovation Program, Grant Agreement No. 101131928. The data analysis has been performed on the HPC workstations "Merac2A" financially supported by the MERAC Foundation and the European Astronomical Society through the 2023 MERAC Prize and the MERAC Support to research (PI Manuel Arca Sedda).

The research leading to these results has been conceived and
developed within the \textit{Einstein} Telescope Observational
Science Board ET-0611A-25.

\end{acknowledgements}


\bibliographystyle{aa_edited}
\bibliography{main}

@article{Reitze:2019iox,
    author = "Reitze, David and others",
    title = "{Cosmic Explorer: The U.S. Contribution to Gravitational-Wave Astronomy beyond LIGO}",
    eprint = "1907.04833",
    archivePrefix = "arXiv",
    primaryClass = "astro-ph.IM",
    reportNumber = "LIGO-P1900316",
    journal = "Bull. Am. Astron. Soc.",
    volume = "51",
    number = "7",
    pages = "035",
    year = "2019"
}

@article{Punturo:2010zz,
    author = "Punturo, M. and others",
    editor = "Ricci, Fulvio",
    title = "{The Einstein Telescope: A third-generation gravitational wave observatory}",
    doi = "10.1088/0264-9381/27/19/194002",
    journal = "Class. Quant. Grav.",
    volume = "27",
    pages = "194002",
    year = "2010"
}

@article{ET:2019dnz,
    doi = {10.1088/1475-7516/2020/03/050},
    url = {https://dx.doi.org/10.1088/1475-7516/2020/03/050},
    year = {2020},
    month = {mar},
    publisher = {},
    volume = {2020},
    number = {03},
    pages = {050},
    author = {Maggiore, Michele and Broeck, Chris Van Den and Bartolo, Nicola and Belgacem, Enis and Bertacca, Daniele and Bizouard, Marie Anne and Branchesi, Marica and Clesse, Sebastien and Foffa, Stefano and García-Bellido, Juan and Grimm, Stefan and Harms, Jan and Hinderer, Tanja and Matarrese, Sabino and Palomba, Cristiano and Peloso, Marco and Ricciardone, Angelo and Sakellariadou, Mairi},
    title = {Science case for the Einstein telescope},
    journal = {Journal of Cosmology and Astroparticle Physics}
}

@article{Branchesi:2023mws,
    author = "Branchesi, Marica and others",
    title = "{Science with the Einstein Telescope: a comparison of different designs}",
    eprint = "2303.15923",
    archivePrefix = "arXiv",
    primaryClass = "gr-qc",
    reportNumber = "ET-0084A-23",
    doi = "10.1088/1475-7516/2023/07/068",
    journal = "JCAP",
    volume = "07",
    pages = "068",
    year = "2023"
}

@article{Ashton:2018jfp,
    author = "Ashton, Gregory and others",
    title = "{BILBY: A user-friendly Bayesian inference library for gravitational-wave astronomy}",
    eprint = "1811.02042",
    archivePrefix = "arXiv",
    primaryClass = "astro-ph.IM",
    doi = "10.3847/1538-4365/ab06fc",
    journal = "Astrophys. J. Suppl.",
    volume = "241",
    number = "2",
    pages = "27",
    year = "2019"
}

@article{Dax:2022pxd,
    author = {Dax, Maximilian and Green, Stephen R. and Gair, Jonathan and P{\"u}rrer, Michael and Wildberger, Jonas and Macke, Jakob H. and Buonanno, Alessandra and Sch{\"o}lkopf, Bernhard},
    title = "{Neural Importance Sampling for Rapid and Reliable Gravitational-Wave Inference}",
    eprint = "2210.05686",
    archivePrefix = "arXiv",
    primaryClass = "gr-qc",
    reportNumber = "LIGO-P2200297",
    doi = "10.1103/PhysRevLett.130.171403",
    journal = "Phys. Rev. Lett.",
    volume = "130",
    number = "17",
    pages = "171403",
    year = "2023"
}

@ARTICLE{2020PhRvD.102j4057G,
       author = {{Green}, Stephen R. and {Simpson}, Christine and {Gair}, Jonathan},
        title = "{Gravitational-wave parameter estimation with autoregressive neural network flows}",
      journal = {\prd},
         year = 2020,
        month = nov,
       volume = {102},
       number = {10},
          eid = {104057},
        pages = {104057},
          doi = {10.1103/PhysRevD.102.104057},
archivePrefix = {arXiv},
       eprint = {2002.07656},
 primaryClass = {astro-ph.IM},
       adsurl = {https://ui.adsabs.harvard.edu/abs/2020PhRvD.102j4057G}
}

@article{Williams:2021qyt,
    author = "Williams, Michael J. and Veitch, John and Messenger, Chris",
    title = "{Nested sampling with normalizing flows for gravitational-wave inference}",
    eprint = "2102.11056",
    archivePrefix = "arXiv",
    primaryClass = "gr-qc",
    doi = "10.1103/PhysRevD.103.103006",
    journal = "Phys. Rev. D",
    volume = "103",
    number = "10",
    pages = "103006",
    year = "2021"
}

@ARTICLE{2023MLS&T...4c5011W,
       author = {{Williams}, Michael J. and {Veitch}, John and {Messenger}, Chris},
        title = "{Importance nested sampling with normalising flows}",
      journal = {Machine Learning: Science and Technology},
         year = 2023,
        month = sep,
       volume = {4},
       number = {3},
          eid = {035011},
        pages = {035011},
          doi = {10.1088/2632-2153/acd5aa},
archivePrefix = {arXiv},
       eprint = {2302.08526},
 primaryClass = {astro-ph.IM},
       adsurl = {https://ui.adsabs.harvard.edu/abs/2023MLS&T...4c5011W}
}

@ARTICLE{2023ApJ...958..129W,
       author = {{Wong}, Kaze W.~K. and {Isi}, Maximiliano and {Edwards}, Thomas D.~P.},
        title = "{Fast Gravitational-wave Parameter Estimation without Compromises}",
      journal = {\apj},
         year = 2023,
        month = dec,
       volume = {958},
       number = {2},
          eid = {129},
        pages = {129},
          doi = {10.3847/1538-4357/acf5cd},
archivePrefix = {arXiv},
       eprint = {2302.05333},
 primaryClass = {astro-ph.IM},
       adsurl = {https://ui.adsabs.harvard.edu/abs/2023ApJ...958..129W}
}

@ARTICLE{2024PhRvD.110h3033W,
       author = {{Wouters}, Thibeau and {Pang}, Peter T.~H. and {Dietrich}, Tim and {Van Den Broeck}, Chris},
        title = "{Robust parameter estimation within minutes on gravitational wave signals from binary neutron star inspirals}",
      journal = {\prd},
         year = 2024,
        month = oct,
       volume = {110},
       number = {8},
          eid = {083033},
        pages = {083033},
          doi = {10.1103/PhysRevD.110.083033},
archivePrefix = {arXiv},
       eprint = {2404.11397},
 primaryClass = {astro-ph.IM},
       adsurl = {https://ui.adsabs.harvard.edu/abs/2024PhRvD.110h3033W}
}

@ARTICLE{2023JOSS....8.5021W,
       author = {{Wong}, Kaze W.~K. and {Gabri{\'e}}, Marylou and {Foreman-Mackey}, Daniel},
        title = "{flowMC: Normalizing flow enhanced sampling package for probabilistic inference in JAX}",
      journal = {The Journal of Open Source Software},
         year = 2023,
        month = mar,
       volume = {8},
       number = {83},
          eid = {5021},
        pages = {5021},
          doi = {10.21105/joss.05021},
archivePrefix = {arXiv},
       eprint = {2211.06397},
 primaryClass = {astro-ph.IM},
       adsurl = {https://ui.adsabs.harvard.edu/abs/2023JOSS....8.5021W}
}

@article{2024arXiv241202651H,
     author = "Hu, Qian and Veitch, John",
     title = "{Costs of Bayesian parameter estimation in third-generation gravitational wave detectors: An assessment of current acceleration methods}",
     eprint = "2412.02651",
     archivePrefix = "arXiv",
     primaryClass = "gr-qc",
     reportNumber = "ET-0666A-24",
     doi = "10.1103/dj7k-tk37",
     journal = "Phys. Rev. D",
     volume = "112",
     number = "8",
     pages = "084039",
     year = "2025"
}

@article{Narola:2024qdh,
    author = "Narola, Harsh and others",
    title = "{Null-stream-based third-generation-ready glitch mitigation for gravitational wave measurements}",
    eprint = "2411.15506",
    archivePrefix = "arXiv",
    primaryClass = "gr-qc",
    doi = "10.1103/l6tp-ykxp",
    journal = "Phys. Rev. D",
    volume = "112",
    number = "2",
    pages = "024079",
    year = "2025"
}

@ARTICLE{2010arXiv1007.4820C,
       author = {{Cornish}, Neil J.},
        title = "{Fast Fisher Matrices and Lazy Likelihoods}",
      journal = {arXiv e-prints},
         year = 2010,
        month = jul,
          eid = {arXiv:1007.4820},
        pages = {arXiv:1007.4820},
          doi = {10.48550/arXiv.1007.4820},
archivePrefix = {arXiv},
       eprint = {1007.4820},
 primaryClass = {gr-qc},
       adsurl = {https://ui.adsabs.harvard.edu/abs/2010arXiv1007.4820C}
}

@article{Veitch:2014wba,
    author = "Veitch, J. and others",
    title = "{Parameter estimation for compact binaries with ground-based gravitational-wave observations using the LALInference software library}",
    eprint = "1409.7215",
    archivePrefix = "arXiv",
    primaryClass = "gr-qc",
    reportNumber = "LIGO-P1400152",
    doi = "10.1103/PhysRevD.91.042003",
    journal = "Phys. Rev. D",
    volume = "91",
    number = "4",
    pages = "042003",
    year = "2015"
}

@ARTICLE{2021PhRvD.104l3030L,
       author = {{Leslie}, Nathaniel and {Dai}, Liang and {Pratten}, Geraint},
        title = "{Mode-by-mode relative binning: Fast likelihood estimation for gravitational waveforms with spin-orbit precession and multiple harmonics}",
      journal = {\prd},
         year = 2021,
        month = dec,
       volume = {104},
       number = {12},
          eid = {123030},
        pages = {123030},
          doi = {10.1103/PhysRevD.104.123030},
archivePrefix = {arXiv},
       eprint = {2109.09872},
 primaryClass = {astro-ph.IM},
       adsurl = {https://ui.adsabs.harvard.edu/abs/2021PhRvD.104l3030L}
}

@ARTICLE{2021PhRvD.104d4062M,
       author = {{Morisaki}, Soichiro},
        title = "{Accelerating parameter estimation of gravitational waves from compact binary coalescence using adaptive frequency resolutions}",
      journal = {\prd},
         year = 2021,
        month = aug,
       volume = {104},
       number = {4},
          eid = {044062},
        pages = {044062},
          doi = {10.1103/PhysRevD.104.044062},
archivePrefix = {arXiv},
       eprint = {2104.07813},
 primaryClass = {gr-qc},
       adsurl = {https://ui.adsabs.harvard.edu/abs/2021PhRvD.104d4062M}
}

@ARTICLE{2015PhRvL.114g1104C,
       author = {{Canizares}, Priscilla and {Field}, Scott E. and {Gair}, Jonathan and {Raymond}, Vivien and {Smith}, Rory and {Tiglio}, Manuel},
        title = "{Accelerated Gravitational Wave Parameter Estimation with Reduced Order Modeling}",
      journal = {\prl},
         year = 2015,
        month = feb,
       volume = {114},
       number = {7},
          eid = {071104},
        pages = {071104},
          doi = {10.1103/PhysRevLett.114.071104},
archivePrefix = {arXiv},
       eprint = {1404.6284},
 primaryClass = {gr-qc},
       adsurl = {https://ui.adsabs.harvard.edu/abs/2015PhRvL.114g1104C}
}

@article{Thrane:2018qnx,
    author = "Thrane, Eric and Talbot, Colm",
    title = "{An introduction to Bayesian inference in gravitational-wave astronomy: parameter estimation, model selection, and hierarchical models}",
    eprint = "1809.02293",
    archivePrefix = "arXiv",
    primaryClass = "astro-ph.IM",
    doi = "10.1017/pasa.2019.2",
    journal = "Publ. Astron. Soc. Austral.",
    volume = "36",
    pages = "e010",
    year = "2019",
    note = "[Erratum: Publ.Astron.Soc.Austral. 37, e036 (2020)]"
}

@ARTICLE{2023PhRvD.107h4037T,
       author = {{Tissino}, Jacopo and {Carullo}, Gregorio and {Breschi}, Matteo and {Gamba}, Rossella and {Schmidt}, Stefano and {Bernuzzi}, Sebastiano},
        title = "{Combining effective-one-body accuracy and reduced-order-quadrature speed for binary neutron star merger parameter estimation with machine learning}",
      journal = {\prd},
         year = 2023,
        month = apr,
       volume = {107},
       number = {8},
          eid = {084037},
        pages = {084037},
          doi = {10.1103/PhysRevD.107.084037},
archivePrefix = {arXiv},
       eprint = {2210.15684},
 primaryClass = {gr-qc},
       adsurl = {https://ui.adsabs.harvard.edu/abs/2023PhRvD.107h4037T}
}

@INPROCEEDINGS{2023mla..confE..34W,
       author = {{Wildberger}, Jonas Bernhard and {Dax}, Maximilian and {Buchholz}, Simon and {Green}, Stephen R. and {Macke}, Jakob and {Sch{\"o}lkopf}, Bernhard},
        title = "{Flow Matching for Scalable Simulation-Based Inference}",
    booktitle = {Machine Learning for Astrophysics},
         year = 2023,
        month = jul,
          eid = {34},
        pages = {34},
          doi = {10.48550/arXiv.2305.17161},
archivePrefix = {arXiv},
       eprint = {2305.17161},
 primaryClass = {cs.LG},
       adsurl = {https://ui.adsabs.harvard.edu/abs/2023mla..confE..34W}
}

@article{Kobyzev_2021,
   title={Normalizing Flows: An Introduction and Review of Current Methods},
   volume={43},
   ISSN={1939-3539},
   url={http://dx.doi.org/10.1109/TPAMI.2020.2992934},
   DOI={10.1109/tpami.2020.2992934},
   number={11},
   journal={IEEE Transactions on Pattern Analysis and Machine Intelligence},
   publisher={Institute of Electrical and Electronics Engineers (IEEE)},
   author={Kobyzev, Ivan and Prince, Simon J.D. and Brubaker, Marcus A.},
   year={2021},
   month=nov, pages={3964–3979} }

@article{Cuoco:2020ogp,
    author = "Cuoco, Elena and others",
    title = "{Enhancing Gravitational-Wave Science with Machine Learning}",
    eprint = "2005.03745",
    archivePrefix = "arXiv",
    primaryClass = "astro-ph.HE",
    doi = "10.1088/2632-2153/abb93a",
    journal = "Mach. Learn. Sci. Tech.",
    volume = "2",
    number = "1",
    pages = "011002",
    year = "2021"
}

@ARTICLE{2023PhRvL.130q1402L,
       author = {{Langendorff}, Jurriaan and {Kolmus}, Alex and {Janquart}, Justin and {Van Den Broeck}, Chris},
        title = "{Normalizing Flows as an Avenue to Studying Overlapping Gravitational Wave Signals}",
      journal = {\prl},
         year = 2023,
        month = apr,
       volume = {130},
       number = {17},
          eid = {171402},
        pages = {171402},
          doi = {10.1103/PhysRevLett.130.171402},
archivePrefix = {arXiv},
       eprint = {2211.15097},
 primaryClass = {gr-qc},
       adsurl = {https://ui.adsabs.harvard.edu/abs/2023PhRvL.130q1402L}
}

@article{Hartwig:2016nde,
    author = "Hartwig, Tilman and Volonteri, Marta and Bromm, Volker and Klessen, Ralf S. and Barausse, Enrico and Magg, Mattis and Stacy, Athena",
    title = "{Gravitational Waves from the Remnants of the First Stars}",
    eprint = "1603.05655",
    archivePrefix = "arXiv",
    primaryClass = "astro-ph.GA",
    doi = "10.1093/mnrasl/slw074",
    journal = "Mon. Not. Roy. Astron. Soc.",
    volume = "460",
    number = "1",
    pages = "L74--L78",
    year = "2016"
}

@article{Belczynski:2016ieo,
    author = "Belczynski, K. and Ryu, T. and Perna, R. and Berti, E. and Tanaka, T. L. and Bulik, T.",
    title = "{On the likelihood of detecting gravitational waves from Population III compact object binaries}",
    eprint = "1612.01524",
    archivePrefix = "arXiv",
    primaryClass = "astro-ph.HE",
    doi = "10.1093/mnras/stx1759",
    journal = "Mon. Not. Roy. Astron. Soc.",
    volume = "471",
    number = "4",
    pages = "4702--4721",
    year = "2017"
}

@article{Gerosa:2017kvu,
    author = "Gerosa, Davide and Berti, Emanuele",
    title = "{Are merging black holes born from stellar collapse or previous mergers?}",
    eprint = "1703.06223",
    archivePrefix = "arXiv",
    primaryClass = "gr-qc",
    doi = "10.1103/PhysRevD.95.124046",
    journal = "Phys. Rev. D",
    volume = "95",
    number = "12",
    pages = "124046",
    year = "2017"
}

@article{Fishbach:2017dwv,
    author = "Fishbach, Maya and Holz, Daniel E. and Farr, Ben",
    title = "{Are LIGO's Black Holes Made From Smaller Black Holes?}",
    eprint = "1703.06869",
    archivePrefix = "arXiv",
    primaryClass = "astro-ph.HE",
    doi = "10.3847/2041-8213/aa7045",
    journal = "Astrophys. J. Lett.",
    volume = "840",
    number = "2",
    pages = "L24",
    year = "2017"
}

@article{Gerosa:2021mno,
    author = "Gerosa, Davide and Fishbach, Maya",
    title = "{Hierarchical mergers of stellar-mass black holes and their gravitational-wave signatures}",
    eprint = "2105.03439",
    archivePrefix = "arXiv",
    primaryClass = "astro-ph.HE",
    doi = "10.1038/s41550-021-01398-w",
    journal = "Nature Astron.",
    volume = "5",
    number = "8",
    pages = "749--760",
    year = "2021"
}

@article{Kritos:2022ggc,
    author = "Kritos, Konstantinos and Strokov, Vladimir and Baibhav, Vishal and Berti, Emanuele",
    title = "{Dynamical formation of black hole binaries in dense star clusters: Rapid cluster evolution code}",
    eprint = "2210.10055",
    archivePrefix = "arXiv",
    primaryClass = "astro-ph.HE",
    doi = "10.1103/PhysRevD.110.043023",
    journal = "Phys. Rev. D",
    volume = "110",
    number = "4",
    pages = "043023",
    year = "2024"
}

@article{kinugawa2020,
	adsurl = {https://ui.adsabs.harvard.edu/abs/2020MNRAS.498.3946K},
	archiveprefix = {arXiv},
	author = {{Kinugawa}, Tomoya and {Nakamura}, Takashi and {Nakano}, Hiroyuki},
	doi = {10.1093/mnras/staa2511},
	eprint = {2005.09795},
	journal = {\mnras},
	month = nov,
	number = {3},
	pages = {3946-3963},
	primaryclass = {astro-ph.HE},
	title = {{Chirp mass and spin of binary black holes from first star remnants}},
	volume = {498},
	year = 2020}

@article{wang2022,
	adsurl = {https://ui.adsabs.harvard.edu/abs/2022MNRAS.515.5106W},
	archiveprefix = {arXiv},
	author = {{Wang}, Long and {Tanikawa}, Ataru and {Fujii}, Michiko},
	doi = {10.1093/mnras/stac2043},
	eprint = {2207.09621},
	journal = {\mnras},
	month = oct,
	number = {4},
	pages = {5106-5120},
	primaryclass = {astro-ph.GA},
	title = {{Gravitational wave of intermediate-mass black holes in Population III star clusters}},
	volume = {515},
	year = 2022}

@article{liubromm2020,
	adsurl = {https://ui.adsabs.harvard.edu/abs/2020MNRAS.495.2475L},
	archiveprefix = {arXiv},
	author = {{Liu}, Boyuan and {Bromm}, Volker},
	doi = {10.1093/mnras/staa1362},
	eprint = {2003.00065},
	journal = {\mnras},
	month = jun,
	number = {2},
	pages = {2475-2495},
	primaryclass = {astro-ph.CO},
	title = {{Gravitational waves from Population III binary black holes formed by dynamical capture}},
	volume = {495},
	year = 2020}

@article{tanikawa2022,
	adsurl = {https://ui.adsabs.harvard.edu/abs/2022PASJ...74..521T},
	archiveprefix = {arXiv},
	author = {{Tanikawa}, Ataru and {Chiaki}, Gen and {Kinugawa}, Tomoya and {Suwa}, Yudai and {Tominaga}, Nozomu},
	doi = {10.1093/pasj/psac010},
	eprint = {2202.00230},
	journal = {\pasj},
	month = jun,
	number = {3},
	pages = {521-532},
	primaryclass = {astro-ph.HE},
	title = {{Can Population III stars be major origins of both merging binary black holes and extremely metal poor stars?}},
	volume = {74},
	year = 2022}

@article{santoliquido2023,
	adsurl = {https://ui.adsabs.harvard.edu/abs/2023MNRAS.524..307S},
	archiveprefix = {arXiv},
	author = {{Santoliquido}, Filippo and {Mapelli}, Michela and {Iorio}, Giuliano and {Costa}, Guglielmo and {Glover}, Simon C.~O. and {Hartwig}, Tilman and {Klessen}, Ralf S. and {Merli}, Lorenzo},
	doi = {10.1093/mnras/stad1860},
	eprint = {2303.15515},
	journal = {\mnras},
	month = sep,
	number = {1},
	pages = {307-324},
	primaryclass = {astro-ph.GA},
	title = {{Binary black hole mergers from population III stars: uncertainties from star formation and binary star properties}},
	volume = {524},
	year = 2023}

@article{DeLuca:2020qqa,
    author = "De Luca, V. and Franciolini, G. and Pani, P. and Riotto, A.",
    title = "{Primordial Black Holes Confront LIGO/Virgo data: Current situation}",
    eprint = "2005.05641",
    archivePrefix = "arXiv",
    primaryClass = "astro-ph.CO",
    doi = "10.1088/1475-7516/2020/06/044",
    journal = "JCAP",
    volume = "06",
    pages = "044",
    year = "2020"
}

@article{Franciolini:2022tfm,
    author = "Franciolini, Gabriele and Musco, Ilia and Pani, Paolo and Urbano, Alfredo",
    title = "{From inflation to black hole mergers and back again: Gravitational-wave data-driven constraints on inflationary scenarios with a first-principle model of primordial black holes across the QCD epoch}",
    eprint = "2209.05959",
    archivePrefix = "arXiv",
    primaryClass = "astro-ph.CO",
    doi = "10.1103/PhysRevD.106.123526",
    journal = "Phys. Rev. D",
    volume = "106",
    number = "12",
    pages = "123526",
    year = "2022"
}

@article{Ng:2022agi,
    author = "Ng, Ken K. Y. and Franciolini, Gabriele and Berti, Emanuele and Pani, Paolo and Riotto, Antonio and Vitale, Salvatore",
    title = "{Constraining High-redshift Stellar-mass Primordial Black Holes with Next-generation Ground-based Gravitational-wave Detectors}",
    eprint = "2204.11864",
    archivePrefix = "arXiv",
    primaryClass = "astro-ph.CO",
    reportNumber = "ET-0068A-22, CE-P2200005",
    doi = "10.3847/2041-8213/ac7aae",
    journal = "Astrophys. J. Lett.",
    volume = "933",
    number = "2",
    pages = "L41",
    year = "2022"
}

@article{PhysRevD.103.104056,
  title = {Computationally efficient models for the dominant and subdominant harmonic modes of precessing binary black holes},
  author = {Pratten, Geraint and Garc\'{\i}a-Quir\'os, Cecilio and Colleoni, Marta and Ramos-Buades, Antoni and Estell\'es, H\'ector and Mateu-Lucena, Maite and Jaume, Rafel and Haney, Maria and Keitel, David and Thompson, Jonathan E. and Husa, Sascha},
  journal = {Phys. Rev. D},
  volume = {103},
  issue = {10},
  pages = {104056},
  numpages = {36},
  year = {2021},
  month = {May},
  publisher = {American Physical Society},
  doi = {10.1103/PhysRevD.103.104056},
  url = {https://link.aps.org/doi/10.1103/PhysRevD.103.104056}
}

@misc{lange2018rapidaccurateparameterinference,
      title={Rapid and accurate parameter inference for coalescing, precessing compact binaries}, 
      author={Jacob Lange and Richard O'Shaughnessy and Monica Rizzo},
      year={2018},
      eprint={1805.10457},
      archivePrefix={arXiv},
      primaryClass={gr-qc},
      url={https://arxiv.org/abs/1805.10457}, 
}

@article{Smith:2021bqc,
    author = "Smith, Rory and others",
    title = "{Bayesian Inference for Gravitational Waves from Binary Neutron Star Mergers in Third Generation Observatories}",
    eprint = "2103.12274",
    archivePrefix = "arXiv",
    primaryClass = "gr-qc",
    reportNumber = "LIGO Document P2100051",
    doi = "10.1103/PhysRevLett.127.081102",
    journal = "Phys. Rev. Lett.",
    volume = "127",
    number = "8",
    pages = "081102",
    year = "2021"
}

@article{Romero-Shaw:2020owr,
    author = "Romero-Shaw, I. M. and others",
    title = "{Bayesian inference for compact binary coalescences with bilby: validation and application to the first LIGO\textendash{}Virgo gravitational-wave transient catalogue}",
    eprint = "2006.00714",
    archivePrefix = "arXiv",
    primaryClass = "astro-ph.IM",
    doi = "10.1093/mnras/staa2850",
    journal = "Mon. Not. Roy. Astron. Soc.",
    volume = "499",
    number = "3",
    pages = "3295--3319",
    year = "2020"
}

@article{Finn:1992wt,
    author = "Finn, Lee S.",
    title = "{Detection, measurement and gravitational radiation}",
    eprint = "gr-qc/9209010",
    archivePrefix = "arXiv",
    reportNumber = "PRINT-93-0128 (NORTHWESTERN)",
    doi = "10.1103/PhysRevD.46.5236",
    journal = "Phys. Rev. D",
    volume = "46",
    pages = "5236--5249",
    year = "1992"
}

@misc{Elvira_2021,
   title={Advances in Importance Sampling},
   ISBN={9781118445112},
   url={http://dx.doi.org/10.1002/9781118445112.stat08284},
   DOI={10.1002/9781118445112.stat08284},
   journal={Wiley StatsRef: Statistics Reference Online},
   publisher={Wiley},
   author={Elvira, Víctor and Martino, Luca},
   year={2021},
   month=feb, pages={1–14} }

@article{https://doi.org/10.1002/wics.56,
    author = {Tokdar, Surya T. and Kass, Robert E.},
    title = {Importance sampling: a review},
    journal = {WIREs Computational Statistics},
    volume = {2},
    number = {1},
    pages = {54-60},
    doi = {https://doi.org/10.1002/wics.56},
    url = {https://wires.onlinelibrary.wiley.com/doi/abs/10.1002/wics.56},
    eprint = {},
    year = {2010}
}

@article{Kong,
    author = {Kong, Arthut},
    title = {A note on importance sampling using standard-
ized weights},
    journal = {University of Chicago, Dept. of Statistics},
    volume = {Tech. Rep 384},
    number = {},
    pages = {},
    doi = {},
    url = {https://d3qi0qp55mx5f5.cloudfront.net/stat/docs/tech-rpts/tr348.pdf},
    eprint = {},
    year = {1992}
}

@book{mcbook,
   author = {Art B. Owen},
   year = 2013,
   title = {Monte Carlo theory, methods and examples},
   publisher = {\url{https://artowen.su.domains/mc/}}
}

@article{Dax:2021myb,
    author = {Dax, Maximilian and Green, Stephen R. and Gair, Jonathan and Deistler, Michael and Sch\"olkopf, Bernhard and Macke, Jakob H.},
    title = "{Group equivariant neural posterior estimation}",
    eprint = "2111.13139",
    archivePrefix = "arXiv",
    journal = "arXiv",
    primaryClass = "cs.LG",
    month = "11",
    year = "2021"
}

@article{Green:2020dnx,
    author = "Green, Stephen R. and Gair, Jonathan",
    title = "{Complete parameter inference for GW150914 using deep learning}",
    eprint = "2008.03312",
    archivePrefix = "arXiv",
    primaryClass = "astro-ph.IM",
    reportNumber = "LIGO-P2000282",
    doi = "10.1088/2632-2153/abfaed",
    journal = "Mach. Learn. Sci. Tech.",
    volume = "2",
    number = "3",
    pages = "03LT01",
    year = "2021"
}

@ARTICLE{61115,
  author={Lin, J.},
  journal={IEEE Transactions on Information Theory}, 
  title={Divergence measures based on the Shannon entropy}, 
  year={1991},
  volume={37},
  number={1},
  pages={145-151},
  doi={10.1109/18.61115}}

@article{10.1214/aoms/1177729694,
author = {S. Kullback and R. A. Leibler},
title = {{On Information and Sufficiency}},
volume = {22},
journal = {The Annals of Mathematical Statistics},
number = {1},
publisher = {Institute of Mathematical Statistics},
pages = {79 -- 86},
year = {1951},
doi = {10.1214/aoms/1177729694},
URL = {https://doi.org/10.1214/aoms/1177729694}
}

@article{Singer:2015ema,
    author = "Singer, Leo P. and Price, Larry R.",
    title = "{Rapid Bayesian position reconstruction for gravitational-wave transients}",
    eprint = "1508.03634",
    archivePrefix = "arXiv",
    primaryClass = "gr-qc",
    reportNumber = "LIGO-P1500009-V3, LIGO-P1500009-V4, LIGO-P1500009-V5, LIGO-P1500009-V6, LIGO-P1500009-V7, LIGO-P1500009-V8",
    doi = "10.1103/PhysRevD.93.024013",
    journal = "Phys. Rev. D",
    volume = "93",
    number = "2",
    pages = "024013",
    year = "2016"
}

@article{Singer:2016eax,
    author = "Singer, Leo P. and others",
    title = "{Going the Distance: Mapping Host Galaxies of LIGO and Virgo Sources in Three Dimensions Using Local Cosmography and Targeted Follow-up}",
    eprint = "1603.07333",
    archivePrefix = "arXiv",
    primaryClass = "astro-ph.HE",
    reportNumber = "LIGO-P1500071-V4, LIGO-P1500071-V5, LIGO-P1500071-V6, LIGO-P1500071-V7",
    doi = "10.3847/2041-8205/829/1/L15",
    journal = "Astrophys. J. Lett.",
    volume = "829",
    number = "1",
    pages = "L15",
    year = "2016"
}

@article{Singer:2016erz,
    author = "Singer, Leo P. and others",
    title = "{Supplement: Going the Distance: Mapping Host Galaxies of LIGO and Virgo Sources in Three Dimensions Using Local Cosmography and Targeted Follow-up}",
    eprint = "1605.04242",
    archivePrefix = "arXiv",
    primaryClass = "astro-ph.IM",
    reportNumber = "LIGO-P1500071-V5, LIGO-P1500071-V6, LIGO-P1500071-V7",
    doi = "10.3847/0067-0049/226/1/10",
    journal = "Astrophys. J. Suppl.",
    volume = "226",
    number = "1",
    pages = "10",
    year = "2016"
}

@article{Marsat:2020rtl,
    author = "Marsat, Sylvain and Baker, John G. and Dal Canton, Tito",
    title = "{Exploring the Bayesian parameter estimation of binary black holes with LISA}",
    eprint = "2003.00357",
    archivePrefix = "arXiv",
    primaryClass = "gr-qc",
    doi = "10.1103/PhysRevD.103.083011",
    journal = "Phys. Rev. D",
    volume = "103",
    number = "8",
    pages = "083011",
    year = "2021"
}

@article{Santoliquido:2024oqs,
    author = "Santoliquido, Filippo and Dupletsa, Ulyana and Tissino, Jacopo and Branchesi, Marica and Iacovelli, Francesco and Iorio, Giuliano and Mapelli, Michela and Gerosa, Davide and Harms, Jan and Pasquato, Mario",
    title = "{Classifying binary black holes from Population III stars with the Einstein Telescope: A machine-learning approach}",
    eprint = "2404.10048",
    archivePrefix = "arXiv",
    primaryClass = "astro-ph.HE",
    doi = "10.1051/0004-6361/202450381",
    journal = "Astron. Astrophys.",
    volume = "690",
    pages = "A362",
    year = "2024"
}

@misc{talts2020validatingbayesianinferencealgorithms,
      title={Validating Bayesian Inference Algorithms with Simulation-Based Calibration}, 
      author={Sean Talts and Michael Betancourt and Daniel Simpson and Aki Vehtari and Andrew Gelman},
      year={2020},
      eprint={1804.06788},
      journal={arXiv},
      archivePrefix={arXiv},
      primaryClass={stat.ME},
      url={https://arxiv.org/abs/1804.06788}, 
}

@article{Cook01092006,
author = {Samantha R Cook, Andrew Gelman and Donald B Rubin},
title = {Validation of Software for Bayesian Models Using Posterior Quantiles},
journal = {Journal of Computational and Graphical Statistics},
volume = {15},
number = {3},
pages = {675--692},
year = {2006},
publisher = {ASA Website},
doi = {10.1198/106186006X136976},


URL = { 
    
        https://doi.org/10.1198/106186006X136976
    
    

},
eprint = { 
    
        https://doi.org/10.1198/106186006X136976
    
    

}

}

@article{Schutz:1986gp,
    author = "Schutz, Bernard F.",
    title = "{Determining the Hubble Constant from Gravitational Wave Observations}",
    doi = "10.1038/323310a0",
    journal = "Nature",
    volume = "323",
    pages = "310--311",
    year = "1986"
}

@article{DelPozzo:2011vcw,
    author = "Del Pozzo, Walter",
    title = "{Inference of the cosmological parameters from gravitational waves: application to second generation interferometers}",
    eprint = "1108.1317",
    archivePrefix = "arXiv",
    primaryClass = "astro-ph.CO",
    doi = "10.1103/PhysRevD.86.043011",
    journal = "Phys. Rev. D",
    volume = "86",
    pages = "043011",
    year = "2012"
}

@article{Chen:2017rfc,
    author = "Chen, Hsin-Yu and Fishbach, Maya and Holz, Daniel E.",
    title = "{A two per cent Hubble constant measurement from standard sirens within five years}",
    eprint = "1712.06531",
    archivePrefix = "arXiv",
    primaryClass = "astro-ph.CO",
    doi = "10.1038/s41586-018-0606-0",
    journal = "Nature",
    volume = "562",
    number = "7728",
    pages = "545--547",
    year = "2018"
}

@article{Libanore:2020fim,
    author = "Libanore, S. and Artale, M. C. and Karagiannis, D. and Liguori, M. and Bartolo, N. and Bouffanais, Y. and Giacobbo, N. and Mapelli, M. and Matarrese, S.",
    title = "{Gravitational Wave mergers as tracers of Large Scale Structures}",
    eprint = "2007.06905",
    archivePrefix = "arXiv",
    primaryClass = "astro-ph.CO",
    doi = "10.1088/1475-7516/2021/02/035",
    journal = "JCAP",
    volume = "02",
    pages = "035",
    year = "2021"
}

@article{Gair:2022zsa,
    author = "Gair, Jonathan R. and others",
    title = "{The Hitchhiker\textquoteright{}s Guide to the Galaxy Catalog Approach for Dark Siren Gravitational-wave Cosmology}",
    eprint = "2212.08694",
    archivePrefix = "arXiv",
    primaryClass = "gr-qc",
    doi = "10.3847/1538-3881/acca78",
    journal = "Astron. J.",
    volume = "166",
    number = "1",
    pages = "22",
    year = "2023"
}

@article{Bosi:2023amu,
    author = "Bosi, M. and Bellomo, N. and Raccanelli, A.",
    title = "{Constraining extended cosmologies with GW\texttimes{}LSS cross-correlations}",
    eprint = "2306.03031",
    archivePrefix = "arXiv",
    primaryClass = "astro-ph.CO",
    reportNumber = "UTWI-20-2023",
    doi = "10.1088/1475-7516/2023/11/086",
    journal = "JCAP",
    volume = "11",
    pages = "086",
    year = "2023"
}

@article{Borghi:2023opd,
    author = "Borghi, Nicola and Mancarella, Michele and Moresco, Michele and Tagliazucchi, Matteo and Iacovelli, Francesco and Cimatti, Andrea and Maggiore, Michele",
    title = "{Cosmology and Astrophysics with Standard Sirens and Galaxy Catalogs in View of Future Gravitational Wave Observations}",
    eprint = "2312.05302",
    archivePrefix = "arXiv",
    primaryClass = "astro-ph.CO",
    doi = "10.3847/1538-4357/ad20eb",
    journal = "Astrophys. J.",
    volume = "964",
    number = "2",
    pages = "191",
    year = "2024"
}

@article{vanderSluys:2009bf,
    author = "van der Sluys, Marc and Mandel, Ilya and Raymond, Vivien and Kalogera, Vicky and Rover, Christian and Christensen, Nelson",
    editor = "Diaz, Mario and Jenet, Fredrick and Mohanty, Soumya",
    title = "{Parameter estimation for signals from compact binary inspirals injected into LIGO data}",
    eprint = "0905.1323",
    archivePrefix = "arXiv",
    primaryClass = "gr-qc",
    doi = "10.1088/0264-9381/26/20/204010",
    journal = "Class. Quant. Grav.",
    volume = "26",
    pages = "204010",
    year = "2009"
}

@article{vanderSluys:2007st,
    author = {van der Sluys, M. V. and R\"over, C. and Stroeer, A. and Raymond, V. and Mandel, I. and Christensen, N. and Kalogera, V. and Meyer, R. and Vecchio, A.},
    title = "{Gravitational-Wave Astronomy with Inspiral Signals of Spinning Compact-Object Binaries}",
    eprint = "0710.1897",
    archivePrefix = "arXiv",
    primaryClass = "astro-ph",
    doi = "10.1086/595279",
    journal = "Astrophys. J. Lett.",
    volume = "688",
    pages = "L61",
    year = "2008"
}

@article{Varma:2021csh,
    author = "Varma, Vijay and Isi, Maximiliano and Biscoveanu, Sylvia and Farr, Will M. and Vitale, Salvatore",
    title = "{Measuring binary black hole orbital-plane spin orientations}",
    eprint = "2107.09692",
    archivePrefix = "arXiv",
    primaryClass = "astro-ph.HE",
    doi = "10.1103/PhysRevD.105.024045",
    journal = "Phys. Rev. D",
    volume = "105",
    number = "2",
    pages = "024045",
    year = "2022"
}

@misc{GWFish_Priors_Tutorial,
  author = {U. Dupletsa},
  title = {{Tutorial on GWFish+Priors}},
  year = {2025},
  url = {https://github.com/janosch314/GWFish/blob/main/priors_tutorial.ipynb},
  note = {Accessed: March 11, 2025}
}

@article{Gorski:1999rt,
    author = "Gorski, Krzysztof M. and Wandelt, Benjamin D. and Hansen, Frode K. and Hivon, Eric and Banday, Anthony J.",
    title = "{The healpix primer}",
    eprint = "astro-ph/9905275",
    archivePrefix = "arXiv",
    journal = "arXiv",
    month = "5",
    year = "1999"
}

@article{Skilling:2004pqw,
    author = "Skilling, John",
    title = "{Nested Sampling}",
    doi = "10.1063/1.1835238",
    journal = "AIP Conf. Proc.",
    volume = "735",
    number = "1",
    pages = "395",
    year = "2004"
}

@software{nessai,
  author       = {Michael J. Williams},
  title        = {nessai: Nested Sampling with Artificial Intelligence},
  month        = feb,
  year         = 2021,
  publisher    = {Zenodo},
  version      = {latest},
  doi          = {10.5281/zenodo.4550693},
  url          = {https://doi.org/10.5281/zenodo.4550693}
}

@ARTICLE{2002ApJ...576..899P,
       author = {{Portegies Zwart}, Simon F. and {McMillan}, Stephen L.~W.},
        title = "{The Runaway Growth of Intermediate-Mass Black Holes in Dense Star Clusters}",
      journal = {\apj},
         year = 2002,
        month = sep,
       volume = {576},
       number = {2},
        pages = {899-907},
          doi = {10.1086/341798},
archivePrefix = {arXiv},
       eprint = {astro-ph/0201055},
 primaryClass = {astro-ph},
       adsurl = {https://ui.adsabs.harvard.edu/abs/2002ApJ...576..899P}
}

@ARTICLE{2023MNRAS.526..429A,
       author = {{Arca Sedda}, Manuel and {Kamlah}, Albrecht W.~H. and {Spurzem}, Rainer and et al.},
        title = "{The DRAGON-II simulations - II. Formation mechanisms, mass, and spin of intermediate-mass black holes in star clusters with up to 1 million stars}",
      journal = {\mnras},
         year = 2023,
        month = nov,
       volume = {526},
       number = {1},
        pages = {429-442},
          doi = {10.1093/mnras/stad2292},
archivePrefix = {arXiv},
       eprint = {2307.04806},
 primaryClass = {astro-ph.GA},
       adsurl = {https://ui.adsabs.harvard.edu/abs/2023MNRAS.526..429A}
}

@ARTICLE{2015MNRAS.454.3150G,
       author = {{Giersz}, Mirek and {Leigh}, Nathan and {Hypki}, Arkadiusz and et al.},
        title = "{MOCCA code for star cluster simulations - IV. A new scenario for intermediate mass black hole formation in globular clusters}",
      journal = {\mnras},
         year = 2015,
        month = dec,
       volume = {454},
       number = {3},
        pages = {3150-3165},
          doi = {10.1093/mnras/stv2162},
archivePrefix = {arXiv},
       eprint = {1506.05234},
 primaryClass = {astro-ph.GA},
       adsurl = {https://ui.adsabs.harvard.edu/abs/2015MNRAS.454.3150G}
}

@ARTICLE{2017MNRAS.467.4180S,
       author = {{Stone}, Nicholas C. and {K{\"u}pper}, Andreas H.~W. and {Ostriker}, Jeremiah P.},
        title = "{Formation of massive black holes in galactic nuclei: runaway tidal encounters}",
      journal = {\mnras},
         year = 2017,
        month = jun,
       volume = {467},
       number = {4},
        pages = {4180-4199},
          doi = {10.1093/mnras/stx097},
archivePrefix = {arXiv},
       eprint = {1606.01909},
 primaryClass = {astro-ph.GA},
       adsurl = {https://ui.adsabs.harvard.edu/abs/2017MNRAS.467.4180S}
}

@ARTICLE{2023MNRAS.521.2930R,
       author = {{Rizzuto}, Francesco Paolo and {Naab}, Thorsten and {Rantala}, Antti and et al.},
        title = "{The growth of intermediate mass black holes through tidal captures and tidal disruption events}",
      journal = {\mnras},
         year = 2023,
        month = may,
       volume = {521},
       number = {2},
        pages = {2930-2948},
          doi = {10.1093/mnras/stad734},
archivePrefix = {arXiv},
       eprint = {2211.13320},
 primaryClass = {astro-ph.GA},
       adsurl = {https://ui.adsabs.harvard.edu/abs/2023MNRAS.521.2930R}
}

@ARTICLE{2019MNRAS.486.5008A,
       author = {{Antonini}, Fabio and {Gieles}, Mark and {Gualandris}, Alessia},
        title = "{Black hole growth through hierarchical black hole mergers in dense star clusters: implications for gravitational wave detections}",
      journal = {\mnras},
         year = 2019,
        month = jul,
       volume = {486},
       number = {4},
        pages = {5008-5021},
          doi = {10.1093/mnras/stz1149},
archivePrefix = {arXiv},
       eprint = {1811.03640},
 primaryClass = {astro-ph.HE},
       adsurl = {https://ui.adsabs.harvard.edu/abs/2019MNRAS.486.5008A}
}

@ARTICLE{2002MNRAS.330..232C,
       author = {{Miller}, M. Coleman and {Hamilton}, Douglas P.},
        title = "{Production of intermediate-mass black holes in globular clusters}",
      journal = {\mnras},
         year = 2002,
        month = feb,
       volume = {330},
       number = {1},
        pages = {232-240},
          doi = {10.1046/j.1365-8711.2002.05112.x},
archivePrefix = {arXiv},
       eprint = {astro-ph/0106188},
 primaryClass = {astro-ph},
       adsurl = {https://ui.adsabs.harvard.edu/abs/2002MNRAS.330..232C}
}

@ARTICLE{2004ApJ...616..221G,
       author = {{G{\"u}ltekin}, Kayhan and {Miller}, M. Coleman and {Hamilton}, Douglas P.},
        title = "{Growth of Intermediate-Mass Black Holes in Globular Clusters}",
      journal = {\apj},
         year = 2004,
        month = nov,
       volume = {616},
       number = {1},
        pages = {221-230},
          doi = {10.1086/424809},
archivePrefix = {arXiv},
       eprint = {astro-ph/0402532},
 primaryClass = {astro-ph},
       adsurl = {https://ui.adsabs.harvard.edu/abs/2004ApJ...616..221G}
}

@ARTICLE{2016MNRAS.459.3432M,
       author = {{Mapelli}, Michela},
        title = "{Massive black hole binaries from runaway collisions: the impact of metallicity}",
      journal = {\mnras},
         year = 2016,
        month = jul,
       volume = {459},
       number = {4},
        pages = {3432-3446},
          doi = {10.1093/mnras/stw869},
archivePrefix = {arXiv},
       eprint = {1604.03559},
 primaryClass = {astro-ph.GA},
       adsurl = {https://ui.adsabs.harvard.edu/abs/2016MNRAS.459.3432M}
}

@ARTICLE{2021A&A...652A..54A,
       author = {{Arca Sedda}, Manuel and {Amaro Seoane}, Pau and {Chen}, Xian},
        title = "{Merging stellar and intermediate-mass black holes in dense clusters: implications for LIGO, LISA, and the next generation of gravitational wave detectors}",
      journal = {\aap},
         year = 2021,
        month = aug,
       volume = {652},
          eid = {A54},
        pages = {A54},
          doi = {10.1051/0004-6361/202037785},
archivePrefix = {arXiv},
       eprint = {2007.13746},
 primaryClass = {astro-ph.GA},
       adsurl = {https://ui.adsabs.harvard.edu/abs/2021A&A...652A..54A}
}

@article{Evans:2023euw,
    author = "Evans, Matthew and others",
    title = "{Cosmic Explorer: A Submission to the NSF MPSAC ngGW Subcommittee}",
    eprint = "2306.13745",
    archivePrefix = "arXiv",
    journal = "arXiv",
    primaryClass = "astro-ph.IM",
    month = "6",
    year = "2023"
}

@article{Gupta:2023lga,
    author = "Gupta, Ish and others",
    title = "{Characterizing gravitational wave detector networks: from A$^\sharp$ to cosmic explorer}",
    eprint = "2307.10421",
    archivePrefix = "arXiv",
    primaryClass = "gr-qc",
    reportNumber = "CE Document No. P2300019, CE Document No. P2300019-v2",
    doi = "10.1088/1361-6382/ad7b99",
    journal = "Class. Quant. Grav.",
    volume = "41",
    number = "24",
    pages = "245001",
    year = "2024"
}

@article{Abac:2025saz,
    author = "Abac, Adrian and others",
    title = "{The Science of the Einstein Telescope}",
    eprint = "2503.12263",
    archivePrefix = "arXiv",
    journal = "arXiv",
    primaryClass = "gr-qc",
    reportNumber = "ET-0036C-25",
    month = "3",
    year = "2025"
}

@article{Vinciguerra:2017ngf,
    author = "Vinciguerra, Serena and Veitch, John and Mandel, Ilya",
    title = "{Accelerating gravitational wave parameter estimation with multi-band template interpolation}",
    eprint = "1703.02062",
    archivePrefix = "arXiv",
    primaryClass = "gr-qc",
    doi = "10.1088/1361-6382/aa6d44",
    journal = "Class. Quant. Grav.",
    volume = "34",
    number = "11",
    pages = "115006",
    year = "2017"
}

@article{Papamakarios:2019fms,
    author = "Papamakarios, George and Nalisnick, Eric and Rezende, Danilo Jimenez and Mohamed, Shakir and Lakshminarayanan, Balaji",
    title = "{Normalizing Flows for Probabilistic Modeling and Inference}",
    eprint = "1912.02762",
    archivePrefix = "arXiv",
    primaryClass = "stat.ML",
    doi = "10.5555/3546258.3546315",
    journal = "J. Machine Learning Res.",
    volume = "22",
    number = "1",
    pages = "2617--2680",
    year = "2021"
}

@article{Aubin:2020goo,
    author = "Aubin, F. and others",
    title = "{The MBTA pipeline for detecting compact binary coalescences in the third LIGO\textendash{}Virgo observing run}",
    eprint = "2012.11512",
    archivePrefix = "arXiv",
    primaryClass = "gr-qc",
    doi = "10.1088/1361-6382/abe913",
    journal = "Class. Quant. Grav.",
    volume = "38",
    number = "9",
    pages = "095004",
    year = "2021"
}

@article{Allene:2025saz,
    author = "All{\'e}n{\'e}, Christopher and others",
    title = "{The MBTA pipeline for detecting compact binary coalescences in the fourth LIGO-Virgo-KAGRA observing run}",
    eprint = "2501.04598",
    archivePrefix = "arXiv",
    primaryClass = "gr-qc",
    doi = "10.1088/1361-6382/add234",
    journal = "Class. Quant. Grav.",
    volume = "42",
    number = "10",
    pages = "105009",
    year = "2025"
}

@article{Fairhurst:2009tc,
    author = "Fairhurst, Stephen",
    title = "{Triangulation of gravitational wave sources with a network of detectors}",
    eprint = "0908.2356",
    archivePrefix = "arXiv",
    primaryClass = "gr-qc",
    doi = "10.1088/1367-2630/11/12/123006",
    journal = "New J. Phys.",
    volume = "11",
    pages = "123006",
    year = "2009",
    note = "[Erratum: New J.Phys. 13, 069602 (2011)]"
}

@article{Baibhav:2019gxm,
    author = "Baibhav, Vishal and Berti, Emanuele and Gerosa, Davide and Mapelli, Michela and Giacobbo, Nicola and Bouffanais, Yann and Di Carlo, Ugo N.",
    title = "{Gravitational-wave detection rates for compact binaries formed in isolation: LIGO/Virgo O3 and beyond}",
    eprint = "1906.04197",
    archivePrefix = "arXiv",
    primaryClass = "gr-qc",
    doi = "10.1103/PhysRevD.100.064060",
    journal = "Phys. Rev. D",
    volume = "100",
    number = "6",
    pages = "064060",
    year = "2019"
}

@article{Mestichelli:2024djn,
    author = "Mestichelli, Benedetta and Mapelli, Michela and Torniamenti, Stefano and Sedda, Manuel Arca and Branchesi, Marica and Costa, Guglielmo and Iorio, Giuliano and Santoliquido, Filippo",
    title = "{Binary black hole mergers from Population III star clusters}",
    eprint = "2405.06037",
    archivePrefix = "arXiv",
    primaryClass = "astro-ph.GA",
    doi = "10.1051/0004-6361/202450667",
    journal = "Astron. Astrophys.",
    volume = "690",
    pages = "A106",
    year = "2024"
}

@article{Christensen:2022bxb,
    author = "Christensen, Nelson and Meyer, Renate",
    title = "{Parameter estimation with gravitational waves}",
    eprint = "2204.04449",
    archivePrefix = "arXiv",
    primaryClass = "gr-qc",
    doi = "10.1103/RevModPhys.94.025001",
    journal = "Rev. Mod. Phys.",
    volume = "94",
    number = "2",
    pages = "025001",
    year = "2022"
}

@article{Singh:2021bwn,
    author = "Singh, Neha and Bulik, Tomasz",
    title = "{Constraining parameters of low mass merging compact binary systems with Einstein Telescope alone}",
    eprint = "2107.11198",
    archivePrefix = "arXiv",
    primaryClass = "astro-ph.HE",
    reportNumber = "Virgo document number VIR-0785A-21",
    doi = "10.1103/PhysRevD.106.123014",
    journal = "Phys. Rev. D",
    volume = "106",
    number = "12",
    pages = "123014",
    year = "2022"
}

@article{Singh:2020wsy,
    author = "Singh, Neha and Bulik, Tomasz",
    title = "{Constraining parameters of coalescing stellar mass binary black hole systems with the Einstein Telescope alone}",
    eprint = "2011.06336",
    archivePrefix = "arXiv",
    primaryClass = "astro-ph.HE",
    reportNumber = "Virgo document number VIR-0876B-20",
    doi = "10.1103/PhysRevD.104.043014",
    journal = "Phys. Rev. D",
    volume = "104",
    number = "4",
    pages = "043014",
    year = "2021"
}

@article{Santoliquido:2025lot,
    author = "Santoliquido, Filippo and others",
    title = "{Fast and accurate parameter estimation of high-redshift sources with the Einstein Telescope}",
    eprint = "2504.21087",
    archivePrefix = "arXiv",
    primaryClass = "astro-ph.HE",
    doi = "10.1103/wf1k-p5cl",
    journal = "Phys. Rev. D",
    volume = "112",
    number = "10",
    pages = "103015",
    year = "2025"
}

@ARTICLE{2022RNAAS...6...89B,
       author = {{Buchner}, Johannes},
        title = "{An Intuition for Physicists: Information Gain From Experiments}",
      journal = {Research Notes of the American Astronomical Society},
         year = 2022,
        month = may,
       volume = {6},
       number = {5},
          eid = {89},
        pages = {89},
          doi = {10.3847/2515-5172/ac6b40},
archivePrefix = {arXiv},
       eprint = {2205.00009},
 primaryClass = {cond-mat.stat-mech},
       adsurl = {https://ui.adsabs.harvard.edu/abs/2022RNAAS...6...89B}
}

@article{Talbot:2023pex,
    author = "Talbot, Colm and Golomb, Jacob",
    title = "{Growing pains: understanding the impact of likelihood uncertainty on hierarchical Bayesian inference for gravitational-wave astronomy}",
    eprint = "2304.06138",
    archivePrefix = "arXiv",
    primaryClass = "astro-ph.IM",
    doi = "10.1093/mnras/stad2968",
    journal = "Mon. Not. Roy. Astron. Soc.",
    volume = "526",
    number = "3",
    pages = "3495--3503",
    year = "2023"
}

@article{Heinzel:2025ogf,
    author = "Heinzel, Jack and Vitale, Salvatore",
    title = "{When (not) to trust Monte Carlo approximations for hierarchical Bayesian inference}",
    eprint = "2509.07221",
    archivePrefix = "arXiv",
    journal = "arXiv",
    primaryClass = "astro-ph.HE",
    month = "9",
    year = "2025"
}

@BOOK{2003itil.book.....M,
       author = {{Mackay}, David J.~C.},
        title = "{Information Theory, Inference and Learning Algorithms}",
         year = 2003,
       adsurl = {https://ui.adsabs.harvard.edu/abs/2003itil.book.....M}
}

@article{Negri:2025cyc,
    author = "Negri, Luca and Samajdar, Anuradha",
    title = "{Neural likelihood estimators for flexible gravitational wave data analysis}",
    eprint = "2509.17606",
    archivePrefix = "arXiv",
    journal = "arXiv",
    primaryClass = "astro-ph.HE",
    month = "9",
    year = "2025"
}

@article{Smith:2019ucc,
    author = "Smith, Rory J. E. and Ashton, Gregory and Vajpeyi, Avi and Talbot, Colm",
    title = "{Massively parallel Bayesian inference for transient gravitational-wave astronomy}",
    eprint = "1909.11873",
    archivePrefix = "arXiv",
    primaryClass = "gr-qc",
    reportNumber = "LIGO Document P1900255-v1",
    doi = "10.1093/mnras/staa2483",
    journal = "Mon. Not. Roy. Astron. Soc.",
    volume = "498",
    number = "3",
    pages = "4492--4502",
    year = "2020"
}

@article{Perret:2025wox,
    author = "Perret, Jules and Ar{\'e}ne, Marc and Porter, Edward K.",
    title = "{DeepHMC : a deep-neural-network acclerated Hamiltonian Monte Carlo algorithm for binary neutron star parameter estimation}",
    eprint = "2505.02589",
    archivePrefix = "arXiv",
    journal = "arXiv",
    primaryClass = "gr-qc",
    month = "5",
    year = "2025"
}

@ARTICLE{2025MNRAS.541..200P,
       author = {{Prathaban}, Metha and {Bevins}, Harry and {Handley}, Will},
        title = "{Accelerated nested sampling with posterior repartitioning and {\ensuremath{\beta}}-flows for gravitational waves}",
      journal = {\mnras},
         year = 2025,
        month = jul,
       volume = {541},
       number = {1},
        pages = {200-213},
          doi = {10.1093/mnras/staf962},
archivePrefix = {arXiv},
       eprint = {2411.17663},
 primaryClass = {astro-ph.IM},
       adsurl = {https://ui.adsabs.harvard.edu/abs/2025MNRAS.541..200P}
}

@article{Prathaban:2025qgg,
    author = "Prathaban, Metha and Yallup, David and Alvey, James and Yang, Ming and Templeton, Will and Handley, Will",
    title = "{Gravitational-wave inference at GPU speed: A bilby-like nested sampling kernel within blackjax-ns}",
    eprint = "2509.04336",
    archivePrefix = "arXiv",
    journal = "arXiv",
    primaryClass = "gr-qc",
    month = "9",
    year = "2025"
}

@article{Zackay:2018qdy,
    author = "Zackay, Barak and Dai, Liang and Venumadhav, Tejaswi",
    title = "{Relative Binning and Fast Likelihood Evaluation for Gravitational Wave Parameter Estimation}",
    eprint = "1806.08792",
    archivePrefix = "arXiv",
    journal = "arXiv",
    primaryClass = "astro-ph.IM",
    month = "6",
    year = "2018"
}

@article{Narola:2023men,
    author = "Narola, Harsh and Janquart, Justin and Meijer, Quirijn and Haris, K. and Van Den Broeck, Chris",
    title = "{Gravitational-wave parameter estimation with relative binning: Inclusion of higher-order modes and precession, and applications to lensing and third-generation detectors}",
    eprint = "2308.12140",
    archivePrefix = "arXiv",
    primaryClass = "gr-qc",
    doi = "10.1103/PhysRevD.110.084085",
    journal = "Phys. Rev. D",
    volume = "110",
    number = "8",
    pages = "084085",
    year = "2024"
}

@article{Morras:2023pug,
    author = "Morras, Gonzalo and Siles, Jose Francisco Nuno and Garcia-Bellido, Juan",
    title = "{Efficient reduced order quadrature construction algorithms for fast gravitational wave inference}",
    eprint = "2307.16610",
    archivePrefix = "arXiv",
    primaryClass = "gr-qc",
    reportNumber = "IFT-UAM/CSIC-23-100",
    doi = "10.1103/PhysRevD.108.123025",
    journal = "Phys. Rev. D",
    volume = "108",
    number = "12",
    pages = "123025",
    year = "2023"
}

@article{Planck:2018vyg,
    author = "Aghanim, N. and others",
    collaboration = "Planck",
    title = "{Planck 2018 results. VI. Cosmological parameters}",
    eprint = "1807.06209",
    archivePrefix = "arXiv",
    primaryClass = "astro-ph.CO",
    doi = "10.1051/0004-6361/201833910",
    journal = "Astron. Astrophys.",
    volume = "641",
    pages = "A6",
    year = "2020",
    note = "[Erratum: Astron.Astrophys. 652, C4 (2021)]"
}

@article{Ashton:2025xba,
    author = "Ashton, Gregory",
    title = "{Reconstructing and resampling: a guide to utilising posterior samples from gravitational wave observations}",
    eprint = "2510.11197",
    archivePrefix = "arXiv",
    journal = "arXiv",
    primaryClass = "gr-qc",
    month = "10",
    year = "2025"
}

@article{Gorski:2004by,
    author = "G{\'o}rski, K. M. and Hivon, E. and Banday, A. J. and Wandelt, B. D. and Hansen, F. K. and Reinecke, M. and Bartelman, M.",
    title = "{HEALPix - A Framework for high resolution discretization, and fast analysis of data distributed on the sphere}",
    eprint = "astro-ph/0409513",
    archivePrefix = "arXiv",
    doi = "10.1086/427976",
    journal = "Astrophys. J.",
    volume = "622",
    pages = "759--771",
    year = "2005"
}

@book{foley1996computer,
  author = {Foley, James D. and van Dam, Andries and Feiner, Steven K. and Hughes, John F.},
  edition = {second},
  isbn = {0201848406},
  publisher = {Addison-Wesley},
  title = {Computer Graphics: Principles and Practice.},
  year = 1996
}

@article{Green:2020hst,
    author = "Green, Stephen R. and Simpson, Christine and Gair, Jonathan",
    title = "{Gravitational-wave parameter estimation with autoregressive neural network flows}",
    eprint = "2002.07656",
    archivePrefix = "arXiv",
    primaryClass = "astro-ph.IM",
    reportNumber = "LIGO-P2000053",
    doi = "10.1103/PhysRevD.102.104057",
    journal = "Phys. Rev. D",
    volume = "102",
    number = "10",
    pages = "104057",
    year = "2020"
}

@article{Dax:2021tsq,
    author = {Dax, Maximilian and Green, Stephen R. and Gair, Jonathan and Macke, Jakob H. and Buonanno, Alessandra and Sch{\"o}lkopf, Bernhard},
    title = "{Real-Time Gravitational Wave Science with Neural Posterior Estimation}",
    eprint = "2106.12594",
    archivePrefix = "arXiv",
    primaryClass = "gr-qc",
    reportNumber = "LIGO-P2100223",
    doi = "10.1103/PhysRevLett.127.241103",
    journal = "Phys. Rev. Lett.",
    volume = "127",
    number = "24",
    pages = "241103",
    year = "2021"
}

@article{Maggiore:2024cwf,
    author = "Maggiore, Michele and Iacovelli, Francesco and Belgacem, Enis and Mancarella, Michele and Muttoni, Niccol{\`o}",
    title = "{Comparison of global networks of third-generation gravitational-wave detectors}",
    eprint = "2411.05754",
    archivePrefix = "arXiv",
    primaryClass = "gr-qc",
    doi = "10.1088/1361-6382/ae110b",
    journal = "Class. Quant. Grav.",
    volume = "42",
    number = "21",
    pages = "215004",
    year = "2025"
}

@article{Mascioli:2025cnw,
    author = "Mascioli, Adriano Frattale and Crescimbeni, Francesco and Pacilio, Costantino and Pani, Paolo and Pannarale, Francesco",
    title = "{Taming systematics in distance and inclination measurements with gravitational waves: Role of the detector network and higher-order modes}",
    eprint = "2504.12473",
    archivePrefix = "arXiv",
    primaryClass = "gr-qc",
    doi = "10.1103/23cf-j34y",
    journal = "Phys. Rev. D",
    volume = "112",
    number = "6",
    pages = "062003",
    year = "2025"
}

@article{deSouza:2023gjv,
    author = "de Souza, Josiel Mendon{\c{c}}a Soares and Sturani, Riccardo",
    title = "{Luminosity distance uncertainties from gravitational wave detections of binary neutron stars by third generation observatories}",
    eprint = "2302.07749",
    archivePrefix = "arXiv",
    primaryClass = "gr-qc",
    doi = "10.1103/PhysRevD.108.043027",
    journal = "Phys. Rev. D",
    volume = "108",
    number = "4",
    pages = "043027",
    year = "2023"
}

@article{Klimenko:2006rh,
    author = "Klimenko, S. and Mohanty, S. and Rakhmanov, Malik and Mitselmakher, Guenakh",
    editor = "Mio, N.",
    title = "{Constraint likelihood method: Generalization for colored noise}",
    doi = "10.1088/1742-6596/32/1/003",
    journal = "J. Phys. Conf. Ser.",
    volume = "32",
    pages = "12--17",
    year = "2006"
}

@article{Usman:2018imj,
    author = "Usman, Samantha A. and Mills, Joseph C. and Fairhurst, Stephen",
    title = "{Constraining the Inclinations of Binary Mergers from Gravitational-wave Observations}",
    eprint = "1809.10727",
    archivePrefix = "arXiv",
    primaryClass = "gr-qc",
    doi = "10.3847/1538-4357/ab0b3e",
    journal = "Astrophys. J.",
    volume = "877",
    number = "2",
    pages = "82",
    year = "2019"
}

@article{LIGOScientific:2025rid,
    author = "Abac, A. G. and others",
    collaboration = "LIGO Scientific, Virgo, KAGRA",
    title = "{GW250114: Testing Hawking{\textquoteright}s Area Law and the Kerr Nature of Black Holes}",
    eprint = "2509.08054",
    archivePrefix = "arXiv",
    primaryClass = "gr-qc",
    reportNumber = "LIGO-P2500421",
    doi = "10.1103/kw5g-d732",
    journal = "Phys. Rev. Lett.",
    volume = "135",
    number = "11",
    pages = "111403",
    year = "2025"
}

@misc{ligo-india,
  author       = "Kandhasamy, Shivaraj and Bose, Sukanta", 
  year = {2020},
  title        = {LIGO India Observatory (LIO) coordinate system for GW analyses},
  howpublished = {\url{https://dcc.ligo.org/public/0167/T2000158/001/LIO_coordinateSystem.pdf}},
}

@misc{ligo-india-asd,
  author       = {{LIGO Collaboration}}, 
  year = {2020},
  title        = {A\# sensitivity curve},
  howpublished = {\url{https://dcc.ligo.org/LIGO-T2300041/public}},
}

@misc{et,
  author       = {{ET Collaboration}},
  year = {2023},
  title        = {ET 10-km and 15-km noise curve},
  howpublished = {\url{https://apps.et-gw.eu/tds/?r=18213}},
}

@article{Unnikrishnan:2013qwa,
    author = "Unnikrishnan, C. S.",
    title = "{IndIGO and LIGO-India: Scope and plans for gravitational wave research and precision metrology in India}",
    eprint = "1510.06059",
    archivePrefix = "arXiv",
    primaryClass = "physics.ins-det",
    doi = "10.1142/S0218271813410101",
    journal = "Int. J. Mod. Phys. D",
    volume = "22",
    pages = "1341010",
    year = "2013"
}

@misc{ce_asd,
  author       = {{CE Collaboration}},
  year = {2022},
  title        = {CE 40-km baseline noise curve},
  howpublished = {\url{https://dcc.cosmicexplorer.org/CE-T2000017/public}},
}

@article{Lanchares:2025kpb,
    author = "Lanchares, Daniel and Freitas, Osvaldo G. and Gonz{\'a}lez-Nuevo, Joaqu{\'\i}n and Font, Jos{\'e} A.",
    title = "{Assessment of normalizing flows for parameter estimation on time-frequency representations of gravitational-wave data}",
    eprint = "2505.08089",
    archivePrefix = "arXiv",
    journal = "arXiv",
    primaryClass = "gr-qc",
    month = "5",
    year = "2025"
}

@article{Marx:2025ioo,
    author = "Marx, Ethan and Chatterjee, Deep and Desai, Malina and Kumar, Ravi and Benoit, William and Sasli, Argyro and Singer, Leo and Coughlin, Michael W. and Harris, Philip and Katsavounidis, Erik",
    title = "{Likelihood-free inference for gravitational-wave data analysis and public alerts}",
    eprint = "2509.22561",
    archivePrefix = "arXiv",
    journal = "arXiv",
    primaryClass = "gr-qc",
    month = "9",
    year = "2025"
}

@article{DeSanti:2024oap,
    author = "De Santi, Federico and Razzano, Massimiliano and Fidecaro, Francesco and Muccillo, Luca and Papalini, Lucia and Patricelli, Barbara",
    title = "{Deep learning to detect gravitational waves from binary close encounters: Fast parameter estimation using normalizing flows}",
    eprint = "2404.12028",
    archivePrefix = "arXiv",
    primaryClass = "gr-qc",
    doi = "10.1103/PhysRevD.109.102004",
    journal = "Phys. Rev. D",
    volume = "109",
    number = "10",
    pages = "102004",
    year = "2024"
}

@article{Payne:2019wmy,
    author = "Payne, Ethan and Talbot, Colm and Thrane, Eric",
    title = "{Higher order gravitational-wave modes with likelihood reweighting}",
    eprint = "1905.05477",
    archivePrefix = "arXiv",
    primaryClass = "astro-ph.IM",
    doi = "10.1103/PhysRevD.100.123017",
    journal = "Phys. Rev. D",
    volume = "100",
    number = "12",
    pages = "123017",
    year = "2019"
}

@article{Chatterjee:2022ggk,
    author = "Chatterjee, Chayan and Kovalam, Manoj and Wen, Linqing and Beveridge, Damon and Diakogiannis, Foivos and Vinsen, Kevin",
    title = "{Rapid Localization of Gravitational Wave Sources from Compact Binary Coalescences Using Deep Learning}",
    eprint = "2207.14522",
    archivePrefix = "arXiv",
    primaryClass = "gr-qc",
    doi = "10.3847/1538-4357/ad08b7",
    journal = "Astrophys. J.",
    volume = "959",
    number = "1",
    pages = "42",
    year = "2023"
}

@article{Srinivasan:2025etu,
    author = "Srinivasan, Rahul and Barausse, Enrico and Korsakova, Natalia and Trotta, Roberto",
    title = "{Simulation-based population inference of LISA{\textquoteright}s Galactic binaries: Bypassing the global fit}",
    eprint = "2506.22543",
    archivePrefix = "arXiv",
    primaryClass = "astro-ph.GA",
    doi = "10.1103/shym-w46f",
    journal = "Phys. Rev. D",
    volume = "112",
    number = "10",
    pages = "103043",
    year = "2025"
}

@article{Romano:2016dpx,
    author = "Romano, Joseph D. and Cornish, Neil J.",
    title = "{Detection methods for stochastic gravitational-wave backgrounds: a unified treatment}",
    eprint = "1608.06889",
    archivePrefix = "arXiv",
    primaryClass = "gr-qc",
    doi = "10.1007/s41114-017-0004-1",
    journal = "Living Rev. Rel.",
    volume = "20",
    number = "1",
    pages = "2",
    year = "2017"
}

@incollection{Rubin1988UsingTS,
  author       = {Donald B. Rubin},
  title        = {Using the {\(SIR\)} Algorithm to Simulate Posterior Distributions},
  booktitle    = {Bayesian Statistics 3},
  editor       = {J. M. Bernardo and M. H. de Groot and D. V. Lindley and A. F. M. Smith},
  publisher    = {Oxford University Press},
  address      = {Oxford, UK},
  year         = {1988},
  pages        = {395--402}
}

@article{fb099965-2cf5-3955-b5e5-077ac9290f24,
 ISSN = {00905364, 21688966},
 URL = {http://www.jstor.org/stable/2241144},
 author = {J. A. Hartigan and P. M. Hartigan},
 journal = {The Annals of Statistics},
 number = {1},
 pages = {70--84},
 publisher = {Institute of Mathematical Statistics},
 title = {The Dip Test of Unimodality},
 urldate = {2025-11-26},
 volume = {13},
 year = {1985}
}

@article{4c99dc66-50c0-3bf4-8c15-3b2f3d6ec1eb,
 ISSN = {00359254, 14679876},
 URL = {http://www.jstor.org/stable/2347485},
 author = {P. M. Hartigan},
 journal = {Journal of the Royal Statistical Society. Series C (Applied Statistics)},
 number = {3},
 pages = {320--325},
 publisher = {[Royal Statistical Society, Oxford University Press]},
 title = {Algorithm AS 217: Computation of the Dip Statistic to Test for Unimodality},
 urldate = {2025-11-26},
 volume = {34},
 year = {1985}
}

@article{Hall:2019xmm,
    author = "Hall, Evan D. and Evans, Matthew",
    title = "{Metrics for next-generation gravitational-wave detectors}",
    eprint = "1902.09485",
    archivePrefix = "arXiv",
    primaryClass = "astro-ph.IM",
    reportNumber = "LIGO-P1800304",
    doi = "10.1088/1361-6382/ab41d6",
    journal = "Class. Quant. Grav.",
    volume = "36",
    number = "22",
    pages = "225002",
    year = "2019"
}

@article{Kofler:2025dux,
    author = {Kofler, Annalena and Dax, Maximilian and Green, Stephen R. and Wildberger, Jonas and Gupte, Nihar and Macke, Jakob H. and Gair, Jonathan and Buonanno, Alessandra and Sch{\"o}lkopf, Bernhard},
    title = "{Flexible Gravitational-Wave Parameter Estimation with Transformers}",
    eprint = "2512.02968",
    archivePrefix = "arXiv",
    primaryClass = "gr-qc",
    reportNumber = "LIGO-P2500728",
    month = "12",
    year = "2025",
    journal = "arXiv",
}

@article{Zeldovich:1967lct,
    author = "Zel'dovich, Ya. B. and Novikov, I. D.",
    title = "{The Hypothesis of Cores Retarded during Expansion and the Hot Cosmological Model}",
    journal = "Sov. Astron.",
    volume = "10",
    pages = "602",
    year = "1967"
}

@article{Hawking:1974rv,
    author = "Hawking, S. W.",
    title = "{Black hole explosions}",
    doi = "10.1038/248030a0",
    journal = "Nature",
    volume = "248",
    pages = "30--31",
    year = "1974"
}

@article{Chapline:1975ojl,
    author = "Chapline, George F.",
    title = "{Cosmological effects of primordial black holes}",
    doi = "10.1038/253251a0",
    journal = "Nature",
    volume = "253",
    number = "5489",
    pages = "251--252",
    year = "1975"
}

@article{Carr:1975qj,
    author = "Carr, Bernard J.",
    title = "{The Primordial black hole mass spectrum}",
    doi = "10.1086/153853",
    journal = "Astrophys. J.",
    volume = "201",
    pages = "1--19",
    year = "1975"
}

@article{Ivanov:1994pa,
    author = "Ivanov, P. and Naselsky, P. and Novikov, I.",
    title = "{Inflation and primordial black holes as dark matter}",
    reportNumber = "NORDITA-94-12-A",
    doi = "10.1103/PhysRevD.50.7173",
    journal = "Phys. Rev. D",
    volume = "50",
    pages = "7173--7178",
    year = "1994"
}

@article{Guth:2012we,
    author = "Guth, Alan H. and Sfakianakis, Evangelos I.",
    title = "{Density Perturbations in Hybrid Inflation Using a Free Field Theory Time-Delay Approach}",
    eprint = "1210.8128",
    archivePrefix = "arXiv",
    primaryClass = "astro-ph.CO",
    reportNumber = "PREPRINT-MIT-CTP-4415",
    month = "10",
    year = "2012",
    journal = "arXiv",
}

@article{Ivanov:1997ia,
    author = "Ivanov, P.",
    title = "{Nonlinear metric perturbations and production of primordial black holes}",
    eprint = "astro-ph/9708224",
    archivePrefix = "arXiv",
    doi = "10.1103/PhysRevD.57.7145",
    journal = "Phys. Rev. D",
    volume = "57",
    pages = "7145--7154",
    year = "1998"
}

@article{Blinnikov:2016bxu,
    author = "Blinnikov, S. and Dolgov, A. and Porayko, N. K. and Postnov, K.",
    title = "{Solving puzzles of GW150914 by primordial black holes}",
    eprint = "1611.00541",
    archivePrefix = "arXiv",
    primaryClass = "astro-ph.HE",
    doi = "10.1088/1475-7516/2016/11/036",
    journal = "JCAP",
    volume = "11",
    pages = "036",
    year = "2016"
}

@article{Carr:2020gox,
    author = "Carr, Bernard and Kohri, Kazunori and Sendouda, Yuuiti and Yokoyama, Jun'ichi",
    title = "{Constraints on primordial black holes}",
    eprint = "2002.12778",
    archivePrefix = "arXiv",
    primaryClass = "astro-ph.CO",
    reportNumber = "RESCEU-03/20; KEK-Cosmo-249; KEK-TH-2199; IPMU20-0024",
    doi = "10.1088/1361-6633/ac1e31",
    journal = "Rept. Prog. Phys.",
    volume = "84",
    number = "11",
    pages = "116902",
    year = "2021"
}

@article{Franciolini:2023opt,
    author = "Franciolini, Gabriele and Iacovelli, Francesco and Mancarella, Michele and Maggiore, Michele and Pani, Paolo and Riotto, Antonio",
    title = "{Searching for primordial black holes with the Einstein Telescope: Impact of design and systematics}",
    eprint = "2304.03160",
    archivePrefix = "arXiv",
    primaryClass = "gr-qc",
    doi = "10.1103/PhysRevD.108.043506",
    journal = "Phys. Rev. D",
    volume = "108",
    number = "4",
    pages = "043506",
    year = "2023"
}

@article{Wouters:2025csq,
    author = "Wouters, Thibeau and Pang, Peter T. H. and Dietrich, Tim and Van Den Broeck, Chris",
    title = "{Incorporating neutron star physics into gravitational wave inference with neural priors}",
    eprint = "2511.22987",
    archivePrefix = "arXiv",
    primaryClass = "astro-ph.HE",
    month = "11",
    year = "2025",
    journal = "arXiv",
}

@article{Chan:2025pdf,
    author = "Chan, Juno C. L. and Maga{\~n}a Zertuche, Lorena and Ezquiaga, Jose Mar{\'\i}a and Lo, Rico K. L. and Vujeva, Luka and Bowman, Joey",
    title = "{Identification and characterization of distorted gravitational waves by lensing using deep learning}",
    eprint = "2511.07186",
    archivePrefix = "arXiv",
    primaryClass = "gr-qc",
    doi = "10.1103/8cz1-kl6n",
    journal = "Phys. Rev. D",
    volume = "113",
    number = "2",
    pages = "024041",
    year = "2026"
}

@article{Colleoni:2024knd,
    author = "Colleoni, Marta and Vidal, Felip A. Ramis and Garc{\'\i}a-Quir{\'o}s, Cecilio and Ak{\c{c}}ay, Sarp and Bera, Sayantani",
    title = "{Fast frequency-domain gravitational waveforms for precessing binaries with a new twist}",
    eprint = "2412.16721",
    archivePrefix = "arXiv",
    primaryClass = "gr-qc",
    doi = "10.1103/PhysRevD.111.104019",
    journal = "Phys. Rev. D",
    volume = "111",
    number = "10",
    pages = "104019",
    year = "2025"
}

@article{Varma:2019csw,
    author = "Varma, Vijay and Field, Scott E. and Scheel, Mark A. and Blackman, Jonathan and Gerosa, Davide and Stein, Leo C. and Kidder, Lawrence E. and Pfeiffer, Harald P.",
    title = "{Surrogate models for precessing binary black hole simulations with unequal masses}",
    eprint = "1905.09300",
    archivePrefix = "arXiv",
    primaryClass = "gr-qc",
    doi = "10.1103/PhysRevResearch.1.033015",
    journal = "Phys. Rev. Research.",
    volume = "1",
    pages = "033015",
    year = "2019"
}

@article{Ramos-Buades:2023yhy,
    author = "Ramos-Buades, Antoni and Buonanno, Alessandra and Gair, Jonathan",
    title = "{Bayesian inference of binary black holes with inspiral-merger-ringdown waveforms using two eccentric parameters}",
    eprint = "2309.15528",
    archivePrefix = "arXiv",
    primaryClass = "gr-qc",
    doi = "10.1103/PhysRevD.108.124063",
    journal = "Phys. Rev. D",
    volume = "108",
    number = "12",
    pages = "124063",
    year = "2023"
}

@article{Thompson:2023ase,
    author = "Thompson, Jonathan E. and Hamilton, Eleanor and London, Lionel and Ghosh, Shrobana and Kolitsidou, Panagiota and Hoy, Charlie and Hannam, Mark",
    title = "{PhenomXO4a: a phenomenological gravitational-wave model for precessing black-hole binaries with higher multipoles and asymmetries}",
    eprint = "2312.10025",
    archivePrefix = "arXiv",
    primaryClass = "gr-qc",
    reportNumber = "LIGO-P2300437",
    doi = "10.1103/PhysRevD.109.063012",
    journal = "Phys. Rev. D",
    volume = "109",
    number = "6",
    pages = "063012",
    year = "2024"
}

\begin{appendix}
\nolinenumbers
\section{Detector and network configurations}
\label{app:detectors}

We provide further details on the detector and network configurations considered in this study. We limited the frequency range from $f_{\mathrm{min}} = 6$ Hz to $f_{\mathrm{max}} = 256$ Hz, using 8-second time segments, which results in frequency bins of size $\mathrm{d}f = 0.125$ Hz. Setting the lower frequency limit to $f_\mathrm{min} = 6$~Hz restricts the duration of all processed signals to less than 8~seconds. This approximation slightly underestimates ET performance, given its expected sensitivity down to $\sim 2$~Hz, but remains adequate for CE, whose sensitivity declines markedly below 5~Hz.

\begin{figure}
    \centering
    \includegraphics[width=1\linewidth]{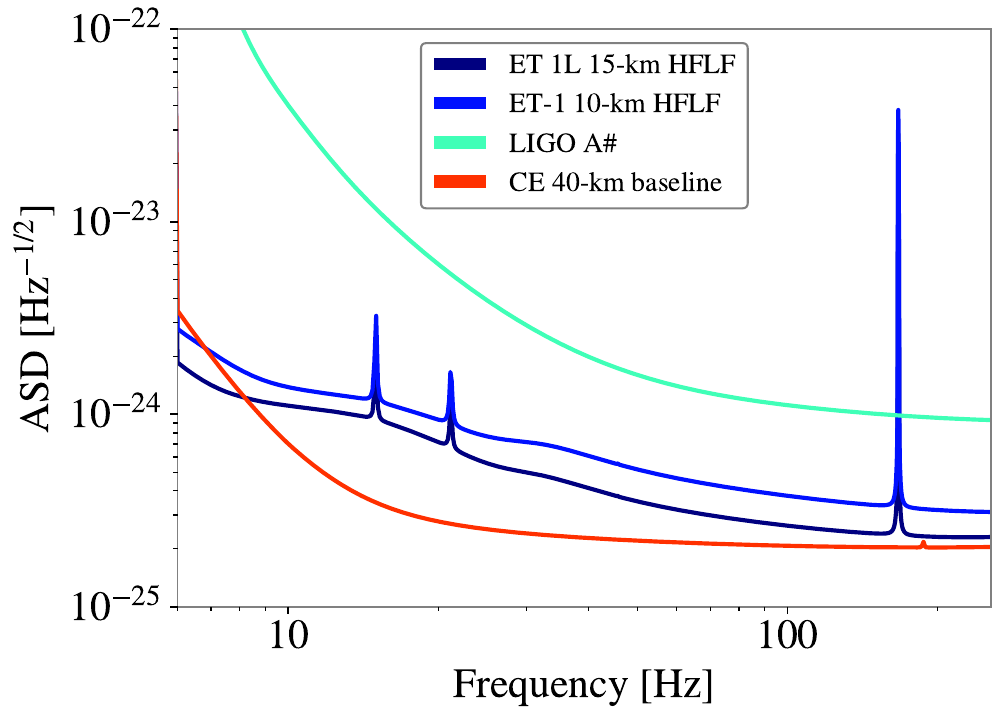}
    \caption{ASDs adopted in this work for different XG detectors. See Sect.~\ref{sec:detectors} and Appendix~\ref{app:detectors} for details.}
    \label{fig:asds}
\end{figure}

All ET configurations ($\Delta$, 2L~A, and 2L~MisA) adopt the high frequency–low frequency (HFLF) cryogenic ASD curve \citep{et}, which incorporates both a high-frequency instrument and a cryogenic low-frequency instrument. Locations and orientations for the ET configurations are taken from \cite{Branchesi:2023mws}. The ASDs for the three LIGO observatories (Livingston, Hanford, and India) at A\# sensitivity is taken from \cite{ligo-india-asd}, while the coordinates and orientation of the Indian interferometer from \cite{ligo-india}. The location 
of CE is assumed to be the near the LIGO-Hanford Observatory (LHO), while the ASD for the 40-km arms in the baseline configuration is taken from \cite{ce_asd}. Figure~\ref{fig:asds} shows the ASDs used in this study for each detector over the specified frequency range.

Table~\ref{tab:detectors} summarizes the locations, orientations, and chord distances between interferometers of the various detector configurations. Chord distances are calculated assuming the Earth is a perfect sphere with radius $R_\oplus = 6378.1366$ km.

\begin{table*}[]
    \caption{Locations (latitude and longitude), orientations ($x$-arm azimuth), and chord distances for the different detector configurations considered in this work.}
    \label{tab:detectors}
    \begin{center}
    \begin{tabular}{llllr}
    \hline\hline
        Detector  & Location &  $x$-arm azimuth & Chord distances & km\\
    \hline
    \multirow{3}{*}{$\Delta$} & ET-1: 40.517°, 9.417°  & ET-1: 250.567°  & ET-1 $\leftrightarrow$ ET-2& 10 \\
             & ET-2: 40.432°, 9.377° & ET-2: 130.567° & ET-2 $\leftrightarrow$ ET-3 & 10\\ 
             & ET-3: 40.448°, 9.493° & ET-3: 10.567° & ET-3 $\leftrightarrow$ ET-1 & 10\\
    \hline
    \multirow{2}{*}{2L A} &   &  ET-S: 0°  &   \multirow{4}{*}{ET-S $\leftrightarrow$ ET-EMR} & \multirow{4}{*}{1166}\\
         & ET-S: 40.517°, 9.417°  & ET-EMR: 0° & &\\     
    \multirow{2}{*}{2L MisA} & ET-EMR: 50.723°, 5.921° & ET-S: 0° &   &\\
            &    & ET-EMR: 45.0° &   &\\
    \hline

    \multirow{2}{*}{Livingston (LLO)} & \multirow{2}{*}{30.563°, $-90.774$°} & \multirow{2}{*}{197.7165°}  & ET-S $\leftrightarrow$ LLO& 7994\\
        &                     &           & ET-EMR $\leftrightarrow$ LLO& 7382 \\
    \hline
    \multirow{3}{*}{Hanford (LHO)} & \multirow{3}{*}{46.455°, $-119.408$°}  & \multirow{3}{*}{125.9994°} &  LLO $\leftrightarrow$ LHO & 3002\\
         &                     &           & ET-S $\leftrightarrow$ LHO& 8352 \\
         &                     &           & ET-EMR $\leftrightarrow$ LHO& 7498 \\
    \hline
    \multirow{4}{*}{India (LIO)} & \multirow{4}{*}{19.613°, 77.031°}  & \multirow{4}{*}{117.6157°} & LIO $\leftrightarrow$ ET-S& 6436 \\
        &                     &           & LIO $\leftrightarrow$ ET-EMR& 6671 \\
        &                     &           & LIO $\leftrightarrow$ LLO& 11488 \\
        &                     &           & LIO $\leftrightarrow$ LHO & 10593 \\
    \hline
    \multirow{3}{*}{CE} & \multirow{3}{*}{46.455°, $-118.592$°}  & \multirow{3}{*}{154.4305°} & CE $\leftrightarrow$ ET-EMR& 7470 \\
       & & & CE $\leftrightarrow$ ET-S & \multirow{2}{*}{8323} \\
       & & & CE $\leftrightarrow$ ET-1 & \\
    \hline
    \end{tabular}
    \end{center}
    \tablefoot{The $x$-arm azimuth is measured counterclockwise from East. For all detectors, the $y$-arm azimuth is given by $x$-arm azimuth + 90°, except for the $\Delta$ configuration, where the $y$-arm azimuth = $x$-arm azimuth + 60°. See Sect.~\ref{sec:detectors} and Appendix~\ref{app:detectors} for details.}
\end{table*}

\section{Waveforms and priors}
\label{app:prior}

We define spin-aligned, quasi-circular waveforms defined by the parameter set
$\theta = \{ \mathcal{M}_\mathrm{d}$, $q$, $D_{\mathrm{L}}$, $\mathrm{dec}$, $\mathrm{ra}$, $\theta_{\mathrm{jn}}$, $\psi$, $t_\mathrm{geocent}$, $\chi_1$, $\chi_2$, $\phi_c \}$,
where $\mathcal{M}_\mathrm{d}$, $q$, and $D_{\mathrm{L}}$ denote the detector-frame chirp mass, mass ratio, and luminosity distance, respectively;
$\mathrm{ra}$ and $\mathrm{dec}$ represent the right ascension and declination of the source; $\theta_{\mathrm{jn}}$ denotes the angle between the line of sight and the total angular momentum ($\vec{J}$). In the absence of precession, as assumed in this work, $\vec{J}$ is equal to the orbital angular momentum of the binary ($\vec{L}$), and thus $\theta_{\mathrm{jn}}$ corresponds to the inclination angle ($\iota$); $\psi$ is the polarization angle; $t_\mathrm{geocent}$ is the coalescence time at the Earth's center; and $\phi_c$ is the phase at coalescence. The waveforms also depend on the aligned-spin parameters $\chi_1$ and $\chi_2$, defined as $\chi_i = a_i \cos \theta_i$, where $a_i = |\chi_i|$ is the spin magnitude and $\theta_i$ is the angle with respect to the orbital angular momentum.

We employ frequency-domain waveforms with a reference frequency fixed at \( f_{\mathrm{ref}} = 20 \)~Hz \citep{Varma:2021csh}, using the {\sc IMRPhenomXPHM} waveform approximant \citep{PhysRevD.103.104056,Colleoni:2024knd}, which includes the subdominant modes \([(\ell, |m|) = (2, 1), (3, 3), (3, 2), (4, 4)]\) in addition to the dominant \([(\ell, |m|) = (2, 2)]\) mode. The prior distributions, $\pi(\theta),$ employed in training {\sc Dingo} are reported in Table~\ref{tab:prior}. In particular, for the detector-frame chirp mass and mass ratio, we chose a prior uniform in primary and secondary masses \citep{GWFish_Priors_Tutorial}; therefore,\begin{equation}
    \label{eq:mchirp}
    \pi(\mathcal{M}_{\mathrm{d}}) \propto \mathcal{M}_{\mathrm{d}},
\end{equation}where $\mathcal{M}_{\mathrm{d}} \in [40,~1100]$ M$_\odot$; and
\begin{equation}
    \label{eq:massratio}
    \pi(q) \propto \frac{(1+q)^{2/5}}{q^{6/5}},
\end{equation}where $q \in [0.125,~1]$. For the luminosity distance, we chose a prior uniform in
comoving volume and source-frame time,
\begin{equation}
\label{eq:dl}
    \pi(D_{\mathrm{L}}) \propto \frac{\mathrm{d} V_c}{\mathrm{d}D_{\mathrm{L}}}\frac{1}{1+z},
\end{equation}where $D_{\mathrm{L}} \in [5,~ 500]$ Gpc, assuming cosmological parameters as provided by \cite{Planck:2018vyg}. For the aligned spin component $\chi_i$, we form a joint prior as follows:
\begin{equation}
\label{eq:spins}
    \pi(\chi_i) = \int_{0}^{0.9}\mathrm{d}a_i  \int_{-1}^{1} \mathrm{d}\cos\theta_i \pi(a_i) \pi(\cos\theta_i) \delta(\chi_i - a_i\cos\theta_i),
\end{equation}where $\pi(a_i) = \mathcal{U}(0,0.9)$ and $\cos\theta_i = \mathcal{U}(-1,1)$ \citep{lange2018rapidaccurateparameterinference,Ashton:2018jfp}.

By convention, we set the Earth's orientation and the positions and orientations of interferometers to those corresponding to the reference time \( t_{\mathrm{ref}} = 1126259462.391 \)~s, which coincides with the GPS trigger time of GW150914 \citep{Romero-Shaw:2020owr}. The merger time (\( t_{\mathrm{geocent}} \)) of each event was randomly sampled around this reference time. Fixing \( t_{\mathrm{ref}} \) does not entail any loss of generality for short signals, as any time offset can be recovered.

\begin{table}
    \caption{Parameters defining GW signals and their adopted priors.}
    \label{tab:prior}
    \centering
    \begin{tabular}{lcr}
    \hline\hline
        Parameter & Units & Prior \\
        \hline
        $\mathcal{M}_{\mathrm{d}}$ & M$_\odot$ & Eq.~\ref{eq:mchirp} \\
        $q$ &  & Eq.~\ref{eq:massratio} \\
        $D_{\mathrm{L}}$ & Gpc & Eq.~\ref{eq:dl} \\
        ra & rad & $\mathcal{U}(0, 2\pi)$ \\
        $\cos$ dec &  & $\mathcal{U}(-1, 1)$ \\
        $\sin \theta_{\mathrm{jn}}$ &  & $\mathcal{U}(0, 1)$ \\
        $\phi_c$ & rad & $\mathcal{U}(0, 2\pi)$ \\
        $\psi$ & rad & $\mathcal{U}(0, \pi)$ \\
        $t_{\mathrm{geocent}}$ & s & $\mathcal{U}(-0.1, +0.1)$\\
        $\chi_i$ &  & Eq.~\ref{eq:spins}  \\
    \hline  
    \end{tabular}
    \tablefoot{$\mathcal{U}(a, b)$ represents a uniform distribution between $a$ and $b$. See Appendix~\ref{app:prior} for details.}
\end{table}

\section{Learning curves and P-P plots}
\label{app:learning}

\begin{figure}
    \centering
    \includegraphics[width=1\linewidth]{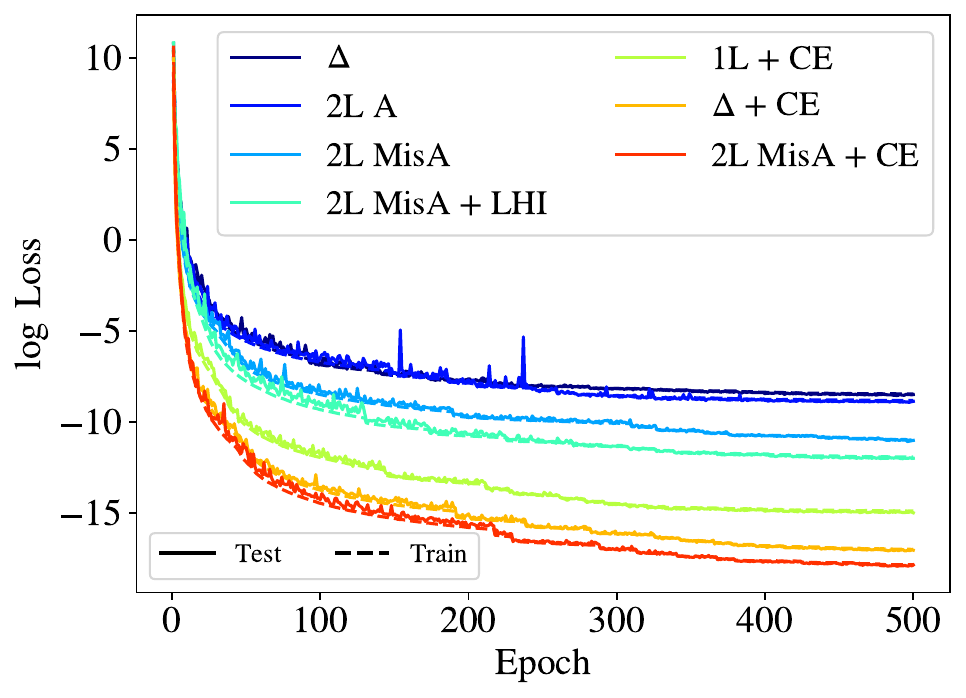}
    \caption{Log loss as a function of training epochs shown for all detector configurations considered in this study. The solid line represents the training set and the dashed line the test set. See Appendix~\ref{app:learning} for further details.}
    \label{fig:loss}
\end{figure}

Figure~\ref{fig:loss} shows that training and test losses remain consistent across all considered detector configurations, indicating the absence of overfitting. We observe an improvement in the training process compared to Fig. 3 of \cite{Santoliquido:2025lot}, where the test losses exhibited significant noise before epoch $\sim 200$.

\begin{figure*}
    \centering
    \includegraphics[width=0.75\linewidth]{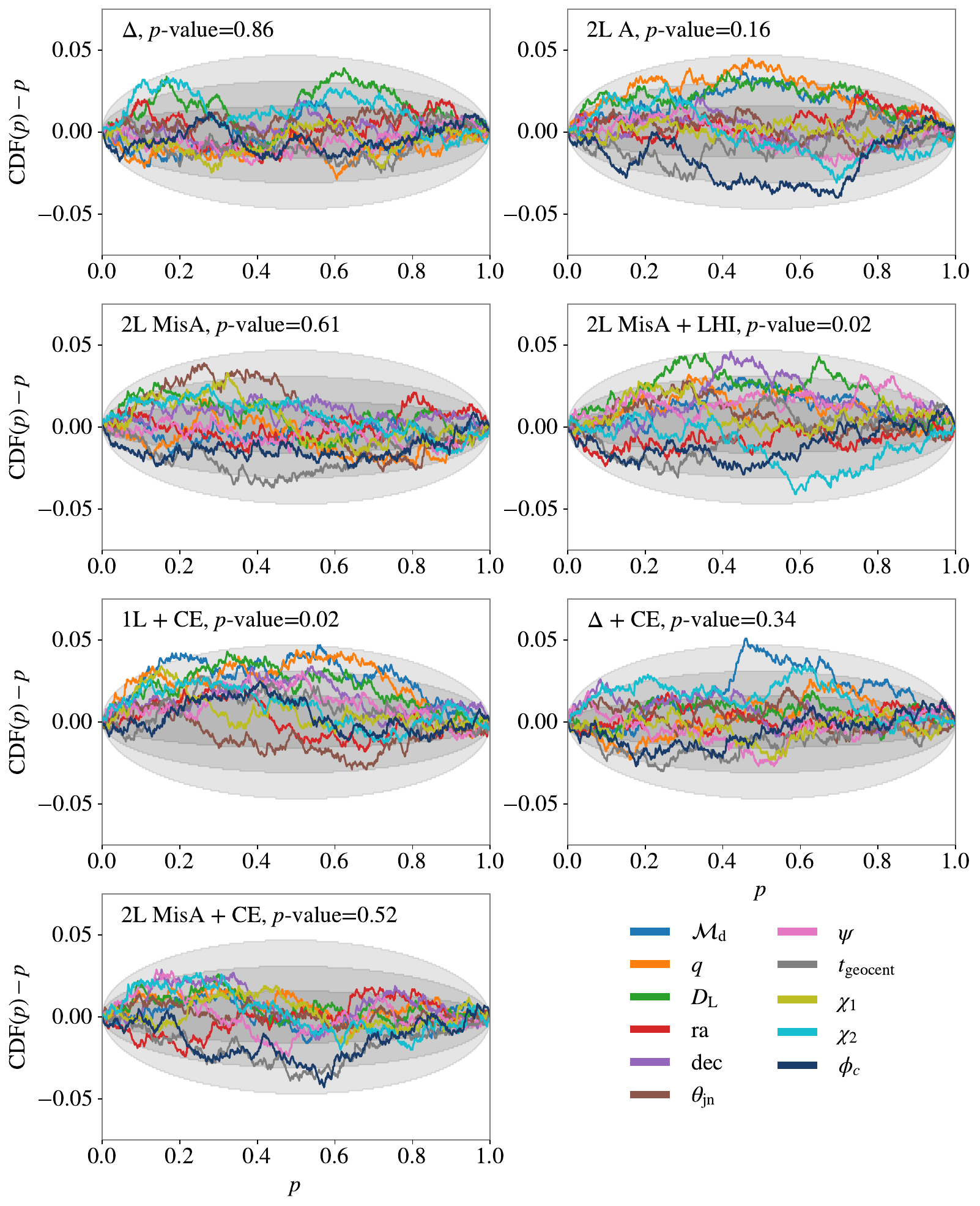}
    \caption{P--P plots showing the confidence interval ($p$) on the $x$-axis versus the difference between the observed fraction of events within that interval and the interval itself ($\mathrm{CDF}(p) - p$) on the $y$-axis, for each GW parameter. Posteriors are obtained with {\sc Dingo} (without importance sampling) from 1000 injected BBH signals sampled from the prior (Table~\ref{tab:prior}) for all the considered detector configurations. Shaded regions denote the 1$\sigma$, 2$\sigma$, and 3$\sigma$ confidence intervals, with combined $p$-values shown in the top-left corner of each panel. Individual $p$-values for each parameter and configuration are reported in Table~\ref{tab:pp}. See Appendix~\ref{app:learning} for further details.}
    \label{fig:ppplots}
\end{figure*}

Figure~\ref{fig:ppplots} displays the probability--probability (P--P) plots for the different detector configurations. P--P plots are commonly used to assess the reliability of inference algorithms \citep{Cook01092006,Veitch:2014wba,talts2020validatingbayesianinferencealgorithms,2020PhRvD.102j4057G}, by verifying whether the fraction of injected parameters recovered within a given credible interval follows a uniform distribution. This method also enables the computation of $p$-values for each parameter through a Kolmogorov--Smirnov test. The results reported in Table~\ref{tab:pp} show good overall consistency. The plots in Fig.~\ref{fig:ppplots} demonstrate that the improved training procedure produces an unbiased proposal distribution $q_\phi(\theta|d)$, as evidenced by the curves lying close to a uniform distribution.

\begin{table*}[]
\caption{$p$-values (color-coded) for each GW parameter (rows) across different detector configurations (columns).}
\label{tab:pp}
\centering
\begin{tabular}{lccccccc}
\hline\hline
  & $\Delta$ & 2L A & 2L MisA & 2L MisA + LHI & 1L + CE & $\Delta$ + CE & 2L MisA + CE \\
\hline
$\mathcal{M}_{\mathrm{d}}$ & \cellcolor{green!43}0.72 & \cellcolor{green!7}0.12 & \cellcolor{green!55}0.92 & \cellcolor{green!17}0.29 & \cellcolor{green!1}0.02 & \cellcolor{green!0}0.01 & \cellcolor{green!54}0.91 \\
$q$ & \cellcolor{green!21}0.35 & \cellcolor{green!1}0.03 & \cellcolor{green!35}0.59 & \cellcolor{green!13}0.23 & \cellcolor{green!2}0.04 & \cellcolor{green!27}0.47 & \cellcolor{green!45}0.76 \\
$D_{\mathrm{L}}$ & \cellcolor{green!4}0.08 & \cellcolor{green!7}0.12 & \cellcolor{green!25}0.42 & \cellcolor{green!1}0.03 & \cellcolor{green!2}0.04 & \cellcolor{green!50}0.84 & \cellcolor{green!28}0.47 \\
ra & \cellcolor{green!47}0.79 & \cellcolor{green!31}0.53 & \cellcolor{green!43}0.73 & \cellcolor{green!35}0.59 & \cellcolor{green!42}0.71 & \cellcolor{green!53}0.90 & \cellcolor{green!34}0.57 \\
dec & \cellcolor{green!48}0.80 & \cellcolor{green!42}0.71 & \cellcolor{green!45}0.76 & \cellcolor{green!1}0.03 & \cellcolor{green!11}0.19 & \cellcolor{green!28}0.47 & \cellcolor{green!20}0.34 \\
$\theta_{\mathrm{jn}}$ & \cellcolor{green!55}0.92 & \cellcolor{green!57}0.96 & \cellcolor{green!5}0.09 & \cellcolor{green!22}0.37 & \cellcolor{green!20}0.34 & \cellcolor{green!43}0.72 & \cellcolor{green!58}0.98 \\
$\psi$ & \cellcolor{green!47}0.79 & \cellcolor{green!43}0.72 & \cellcolor{green!55}0.92 & \cellcolor{green!13}0.23 & \cellcolor{green!15}0.25 & \cellcolor{green!25}0.43 & \cellcolor{green!20}0.35 \\
$t_{\mathrm{geocent}}$ & \cellcolor{green!32}0.55 & \cellcolor{green!18}0.30 & \cellcolor{green!6}0.12 & \cellcolor{green!16}0.27 & \cellcolor{green!21}0.37 & \cellcolor{green!17}0.29 & \cellcolor{green!4}0.08 \\
$\chi_1$ & \cellcolor{green!33}0.55 & \cellcolor{green!60}1.00 & \cellcolor{green!12}0.21 & \cellcolor{green!24}0.41 & \cellcolor{green!11}0.19 & \cellcolor{green!36}0.61 & \cellcolor{green!49}0.83 \\
$\chi_2$ & \cellcolor{green!12}0.21 & \cellcolor{green!16}0.28 & \cellcolor{green!27}0.47 & \cellcolor{green!3}0.06 & \cellcolor{green!25}0.42 & \cellcolor{green!9}0.15 & \cellcolor{green!25}0.43 \\
$\phi_c$ & \cellcolor{green!53}0.89 & \cellcolor{green!4}0.07 & \cellcolor{green!30}0.50 & \cellcolor{green!19}0.33 & \cellcolor{green!30}0.50 & \cellcolor{green!26}0.44 & \cellcolor{green!2}0.04 \\
\hline
\end{tabular}
\tablefoot{See Fig.~\ref{fig:ppplots} and Appendix~\ref{app:learning} for details.}
\end{table*}

\section{{\sc{Dingo-IS}} versus {\sc{Bilby}}}
\label{app:bilby}

\begin{table*}[htbp]
\caption{Injected parameters ($\theta^{\mathrm{inj}}$) of the event shown in Fig.~\ref{fig:dingo_bilby_2L_short}, along with the median JSD ($\langle \mathrm{JSD} \rangle$, in units of $10^{-4}$ nat), which quantifies the deviation between {\sc Dingo-IS} and {\sc Bilby} for one-dimensional marginal posteriors.}
\label{tab:jsd}
\centering
\begin{tabular}{lc|ccc|ccc}
\hline\hline
 & & \multicolumn{3}{c|}{2L MisA} & \multicolumn{3}{c}{2L MisA + CE}\\
Parameters & $\theta^{\mathrm{inj}}$ & $\langle \mathrm{JSD} \rangle$ & JSD$^{\mathrm{thr}}$ & $\theta$ & $\langle \mathrm{JSD} \rangle$ & JSD$^{\mathrm{thr}}$ & $\theta$ \\
\hline
$\mathcal{M}_{\mathrm{d}}/$M$_\odot$  & $749$ & 1.8 & 4.7 & ${757}_{-7}^{+7}$ & 2.9 & 5.5 & ${751}_{-3}^{+3}$ \\
$q$ & $0.73$ & 1.0 & 5.0  & ${0.75}_{-0.02}^{+0.02}$ & 2.0 & 4.7 & ${0.73}_{-0.01}^{+0.01}$ \\
$D_{\mathrm{L}}$/Gpc & $13.8$ & 1.7 & 4.9  & ${18}_{-5}^{+1}$ & 2.2 & 4 & ${13.7}_{-0.3}^{+0.3}$ \\
ra/rad & $4.60$ & 1.7 & 7.3  & ${2.40}_{-0.15}^{+2.25}$ & 0.0 & 0.0 & ${4.60}_{-0.01}^{+0.01}$ \\
dec/rad & $-0.40$ & 1.8 & 5.6  & ${0.16}_{-0.56}^{+0.12}$ & 0.0 & 0.1 & ${-0.40}_{-0.01}^{+0.01}$ \\
${\theta_{\mathrm{jn}}}$/rad & $1.06$  & 0.4 & 2.0  & ${2.04}_{-0.98}^{+0.05}$  & 0.1 & 0.8 & ${1.07}_{-0.01}^{+0.01}$ \\
${\phi_{\mathrm{c}}}$/rad & $0.09$ & 0.5 & 2.1  & ${0.11}_{-0.10}^{+6.17}$ & 7.6 & 3.3 & ${0.08}_{-0.06}^{+0.08}$ \\
${\psi}$/rad & $2.14$ & 2.3 & 7.8  & ${1.62}_{-0.10}^{+0.54}$  & 0.0 & 0.7 & ${2.13}_{-0.01}^{+0.01}$ \\
$t_{\mathrm{geocent}}$/s & $0.056$  & 1.6 & 5.2  & ${0.082}_{-0.029}^{+0.003}$  & 2.9 & 6.6 & ${0.056}_{-0.001}^{+0.001}$ \\
${\chi_1}$ & $0.35$ & 0.7 & 2.6  & ${0.36}_{-0.07}^{+0.07}$ & 2.3 & 3.9 & ${0.36}_{-0.04}^{+0.04}$ \\
${\chi_2}$ & $0.20$ & 1.1 & 3.6  & ${0.22}_{-0.10}^{+0.08}$ & 2.4 & 6.6 & ${0.21}_{-0.06}^{+0.06}$ \\
      
 & & & &  & & & \\
$\rho$ & & & & 232 & & & 331 \\
$m_{1,\mathrm{s}}$/M$_\odot$ & 364.2  & & & ${317}_{-8}^{+66}$ & & & ${366}_{-3}^{+3}$ \\
$m_{2,\mathrm{s}}$/M$_\odot$ & 265.2  & & & ${236}_{-6}^{+49}$ & & & ${267}_{-3}^{+3}$ \\
$z$ & 1.8 & & & ${2.18}_{-0.54}^{+0.09}$ & & & ${1.77}_{-0.03}^{+0.03}$ \\
${\Delta\Omega}_{90\%}$/deg$^2$ & & & & 235 & & & 1 \\
${\Delta V}^{c}_{90\%}$/Mpc$^3$ & & & & $4.1 \times 10^{8}$ & & & $7.2 \times 10^{5}$ \\

& & & &  & & & \\
Sample efficiency/\% & & & & 16 & & & 10 \\
{\sc{Dingo-IS}}  $\log \mathcal{Z}(d) \pm \sigma$ & &  & & $-4046.006 \pm 0.007$ & & & $-6180.05 \pm 0.01$ \\
{\sc{Bilby}}     $\log \mathcal{Z}(d) \pm \sigma$ & &  & & $-4046.29 \pm 0.07$ & & & $-6179.26 \pm 0.07$ \\
Computing time/hours & & & & 51.5   & & & 8.8 \\
\hline
\end{tabular}
\tablefoot{For reference, the corresponding JSD threshold (JSD$^{\mathrm{thr}}$, in units of $10^{-4}$ nat) is also shown. The median and 90\% credible interval of the recovered parameters $\theta$ with {\sc Dingo-IS} are reported for two different detector configurations: 2L~MisA and 2L~MisA~+~CE. The bottom rows display the optimal S/N ($\rho$, Eq.~\ref{eq:snr}), source-frame primary ($m_{1,\mathrm{s}}$) and secondary ($m_{2,\mathrm{s}}$) masses, redshift ($z$), sky and comoving volume localization error ($\Delta \Omega_{90\%}$ and $\Delta V^c_{90\%}$), sample efficiency (Eq.~\ref{eq:eff}), log evidence and its uncertainty computed with {\sc Dingo-IS} (Eq.~\ref{eq:evidence}) and {\sc Bilby}, and computing time in {\sc Bilby} obtained using \texttt{npool = 8} parallel processes. See Appendix~\ref{app:bilby} for details.}
\end{table*}

\begin{figure*}
    \centering
    \includegraphics[width=0.9\linewidth]{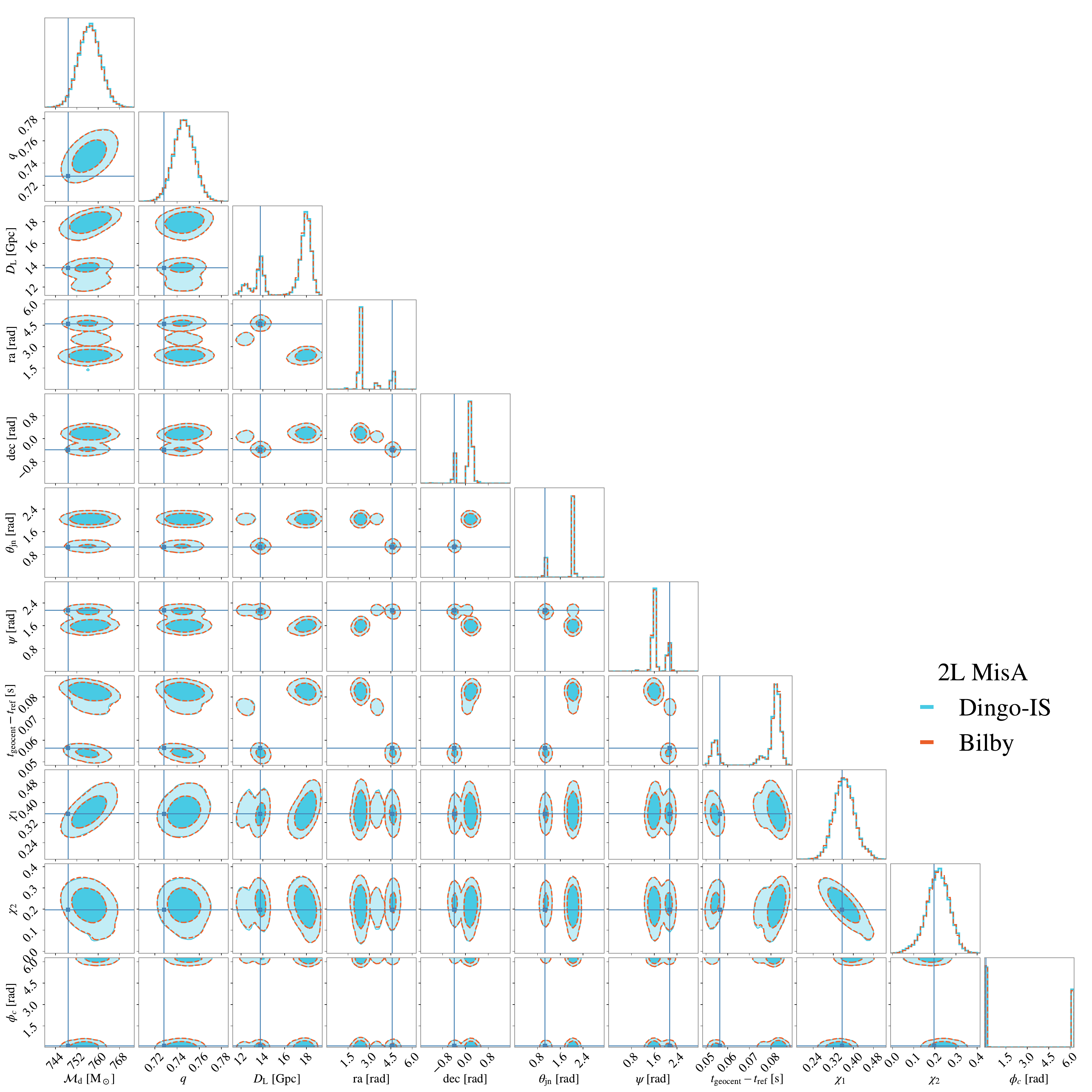}
    \caption{Marginalized one- and two-dimensional posterior distributions for all parameters for the event shown in Fig.~\ref{fig:dingo_bilby_2L_short} observed with the 2L~MisA configuration. Results from {\sc Dingo-IS} (light blue) are compared with {\sc Bilby} (orange). Vertical and horizontal lines indicate the true injected values (see Table~\ref{tab:jsd}) and contours represent the 68\% and 95\% credible regions. See Appendix~\ref{app:bilby} for further details.}
    \label{fig:dingo_bilby_2L}
\end{figure*}

\begin{figure*}
    \centering
    \includegraphics[width=0.9\linewidth]{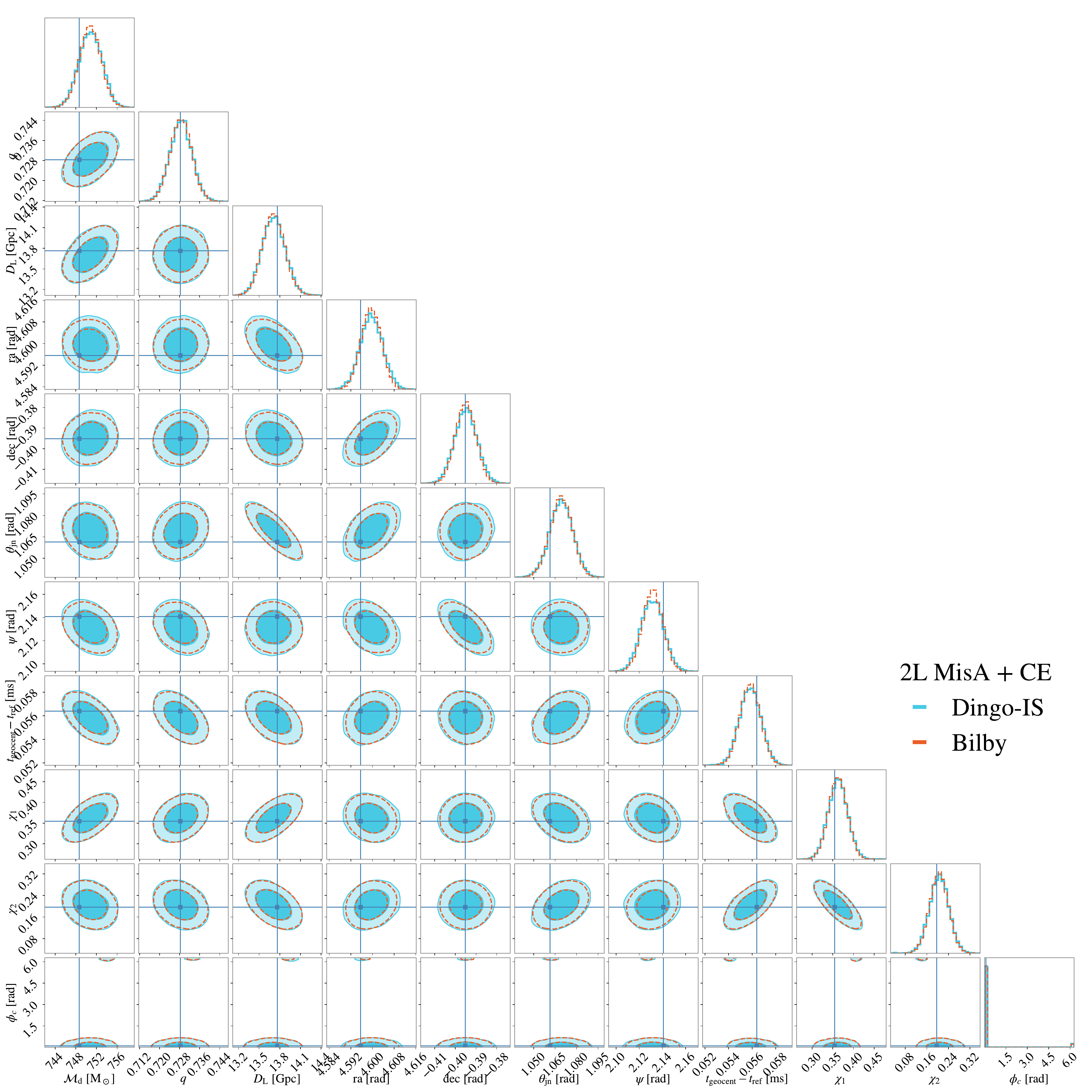}
    \caption{Marginalized one- and two-dimensional posterior distributions for all parameters of the same event shown in Fig.~\ref{fig:dingo_bilby_2L}, this time observed with the 2L~MisA~+~CE configuration. Results from {\sc Dingo-IS} (light blue) are compared with {\sc Bilby} (orange). Vertical and horizontal lines indicate the true injected values (see Table~\ref{tab:jsd}) and contours represent the 68\% and 95\% credible regions. See Appendix~\ref{app:bilby} for further details.}
    \label{fig:dingo_bilby_2L_CE}
\end{figure*}

The parameters ($\theta$) of a GW signal are inferred from strain data ($d$) through Bayes' theorem \citep{vanderSluys:2007st,vanderSluys:2009bf,Thrane:2018qnx,Christensen:2022bxb}:  
\begin{equation}  
\label{eq:bayes}
    p(\theta|d) = \frac{\mathcal{L}(d|\theta)\pi(\theta)}{\mathcal{Z}(d)},  
\end{equation}where $\mathcal{L}(d|\theta)$ is the likelihood function, $\pi(\theta)$ is the prior distribution (see Appendix~\ref{app:prior}) and $\mathcal{Z}(d) = \int \mathrm{d}\theta \mathcal{L}(d|\theta)\pi(\theta)$ is the evidence. Assuming stationary Gaussian noise, the likelihood in Eq.~\ref{eq:bayes} takes the form \citep{Finn:1992wt,Veitch:2014wba,Romano:2016dpx,Ashton:2018jfp}\begin{equation}
\label{eq:likelihood}
    \mathcal{L}(d|\theta) \propto \exp \left[\sum_i \left( (d_i|h_i(\theta)) -\frac{1}{2} (h_i(\theta)\,|\, h_i(\theta))\right) \right],
\end{equation}where the index $i$ labels the detectors in the network, $h_i(\theta)$ denotes the GW signal, and $(\cdot|\cdot)$ the noise-weighted inner product, defined as
\begin{equation}
\label{eq:inner}
    (a|b) = 4\, \mathrm{Re} \int_{f_{\mathrm{min}}}^{f_{\mathrm{max}}} \mathrm{d}f\, \frac{a^{*}(f)\, b(f)}{S(f)} ,
\end{equation}with $\mathrm{Re}$ indicating the real part and $^{*}$ the complex conjugate. The quantity $\sqrt{S(f)}$ represents the noise ASD (see Appendix~\ref{app:detectors} for details). Equation~\ref{eq:inner} is also used to compute the optimal S/N:
\begin{equation}
\label{eq:snr}
    \rho = \sqrt{(h(\theta^{\mathrm{inj}})|h(\theta^{\mathrm{inj}}))},
\end{equation}where $h(\theta^{\mathrm{inj}})$ denotes the GW signal evaluated with injected parameters. For a network of detectors, the total optimal S/N is obtained by adding the individual detector S/Ns in quadrature.

To sample from Eq. \ref{eq:bayes}, we employed {\sc Bilby} \citep{Romero-Shaw:2020owr}, a Python library designed for Bayesian inference with the nested sampling algorithm {\sc nessai} \citep{nessai}. Thanks to its use of normalizing flows, {\sc nessai} achieves rapid convergence. We fixed the number of live points to \texttt{n\_live = 10\_000}.

For the source shown in Fig.~\ref{fig:dingo_bilby_2L_short} and observed with the 2L MisA configuration, the default {\sc nessai} settings yield suboptimal performance. To improve robustness, we increased the \texttt{volume\_fraction} from 0.95 to 0.98. This parameter controls the fraction of the total probability enclosed by the contour in the latent space---the base space from which samples are initially drawn before being mapped to physical space through the normalizing flow \citep{Williams:2021qyt}. Although a higher volume fraction improves performance, it also increases computational cost, which explains the long runtime ($\sim$ 50 hours) required for this source (see Table~\ref{tab:jsd}).

To quantify the agreement between posterior samples obtained with {\sc Dingo-IS} and {\sc Bilby}, we employ the Jensen--Shannon divergence \citep[JSD;][]{61115}, a symmetrized variant of the KL divergence \citep{10.1214/aoms/1177729694}. The JSD provides a measure of the similarity between two probability distributions, taking values from 0~nat, when the distributions are identical, up to $\ln(2) = 0.69$~nat for completely distinct distributions.

To determine a threshold for the JSD beyond which one-dimensional posteriors can be regarded as statistically different, we follow the procedure outlined in \cite{Santoliquido:2025lot}. Using {\sc Dingo-IS}, we generate 100 independent sets of $10^5$ posterior samples. For each parameter $\theta$, we compute the JSD across all 4950 unique pairs of these sets. Because all samples originate from the same underlying distribution, the resulting JSD values capture statistical fluctuations rather than meaningful discrepancies. We adopt the 95th percentile of these values as the threshold for each parameter (JSD$^{\mathrm{thr}}$ in Table~\ref{tab:jsd}). Finally, we compare the one-dimensional posteriors obtained with {\sc Bilby} to each {\sc Dingo-IS} set, and verify whether the median JSD lies below the corresponding threshold ($\langle \mathrm{JSD} \rangle$ in Table~\ref{tab:jsd}).

Table~\ref{tab:jsd} lists the JSD values computed for the marginal one-dimensional posteriors of the source analyzed in Fig.~\ref{fig:dingo_bilby_2L_short}. The results show that the posteriors obtained with {\sc Bilby} and {\sc Dingo-IS} are statistically indistinguishable for both the 2L~MisA and 2L~MisA~+~CE configurations. This excellent agreement is further illustrated in Figs.~\ref{fig:dingo_bilby_2L}~and~\ref{fig:dingo_bilby_2L_CE}, which display the marginalized one- and two-dimensional posteriors for all parameters for the 2L~MisA and 2L~MisA~+~CE networks, respectively.

Table~\ref{tab:jsd} also reports the Bayesian evidence $\mathcal{Z}(d)$ defined in Eq.~\ref{eq:bayes}, along with its uncertainty estimated by averaging the weights (Eq.~\ref{eq:weight}) obtained with {\sc Dingo-IS} \citep{mcbook}:
\begin{equation}
    \label{eq:evidence}
    \mathcal{Z}(d) \pm \sigma = \frac{1}{n} \sum_i w_i \left( 1 \pm \sqrt{\frac{1 - \epsilon}{n - \epsilon}} \right),
\end{equation}where $n$ denotes the total number of samples drawn from the proposal distribution $q_\phi(\theta|d)$, and $\epsilon$ is the sampling efficiency defined in Eq.~\ref{eq:eff}. The log evidences computed with {\sc Dingo-IS} and {\sc Bilby} are consistent within statistical uncertainties.

\section{Information gain}
\label{app:info}

We computed Eq.~\ref{eq:info_gain} through a Monte Carlo integration:
\begin{equation}
I = \int p(\theta|d) \log_2 \frac{p(\theta|d)}{\pi(\theta)}  \mathrm{d}\theta \approx \frac{1}{N_s} \sum_{i=1}^{N_s} \log_2 \frac{p(\theta_i|d)}{\pi(\theta_i)},
\end{equation}where $N_s = 3\times10^4$ denotes the number of samples drawn from the posterior. In our case, these samples were obtained through sampling-importance resampling \citep[][]{Rubin1988UsingTS} based on the importance weights defined in Eq. \ref{eq:weight}. Using Bayes' theorem, we can rewrite the ratio inside the logarithm as
\begin{align}
\label{eq:info}
    I = \frac{1}{N_s} \sum_{i}^{N_s} \log_2 \frac{p(\theta_i|d)}{\pi(\theta_i)} = \frac{1}{N_s} \sum_{i}^{N_s}  \log_2 \frac{\mathcal{L}(d|\theta_i)\cancel{\pi(\theta)}}{\mathcal{Z}(d)\cancel{\pi(\theta)}} = \\ - \log_2 \mathcal{Z}(d) + \frac{1}{N_s} \sum_{i}^{N_s}  \log_2 \mathcal{L}(d|\theta_i),
\end{align}where the log evidence $\log_2 \mathcal{Z}(d)$ is evaluated through the importance sampling weights (Eq.~\ref{eq:evidence}). We estimated the variance associated with the Monte Carlo estimator of the information gain \citep{2003itil.book.....M, Talbot:2023pex, Heinzel:2025ogf} as follows:
\begin{equation}
    \sigma_I^2 = \frac{1}{N_s} \left[ \frac{1}{N_s} \sum_{i=1}^{N_s} \left(-\log_2 \mathcal{Z}(d) + \log_2 \mathcal{L}(d|\theta_i)\right)^2 - I^2 \right],
\end{equation}where $I$ is defined in Eq.~\ref{eq:info} and $\sigma_I^2 < 10^{-3}$~bits for all detector configurations.

\section{Multimodal luminosity distance}
\label{app:multidl}

\begin{figure}
    \centering
    \includegraphics[width=1\linewidth]{unimodal_dL.pdf}
    \caption{Percentage of luminosity distance posteriors ($y$-axis) that are multimodal (dark blue) or unimodal (light blue) for all detector configurations ($x$-axis). See Appendix~\ref{app:multidl} for details.}
    \label{fig:diptest}
\end{figure}

We apply Hartigan’s dip test for unimodality \citep{fb099965-2cf5-3955-b5e5-077ac9290f24,4c99dc66-50c0-3bf4-8c15-3b2f3d6ec1eb} to the one-dimensional marginal posterior of the luminosity distance. A distribution is classified as unimodal if the $p$-value exceeds 0.01. As shown in Fig.~\ref{fig:diptest}, about 19\% (23\%) of sources have a multimodal $D_{\mathrm{L}}$ posterior for 2L~MisA (2L~A), while the $\Delta$ configurations display only $\sim2$\%.

\section{Robustness to waveform approximant variations}
\label{app:waveforms}

\begin{figure*}
    \centering
    \includegraphics[width=0.75\linewidth]{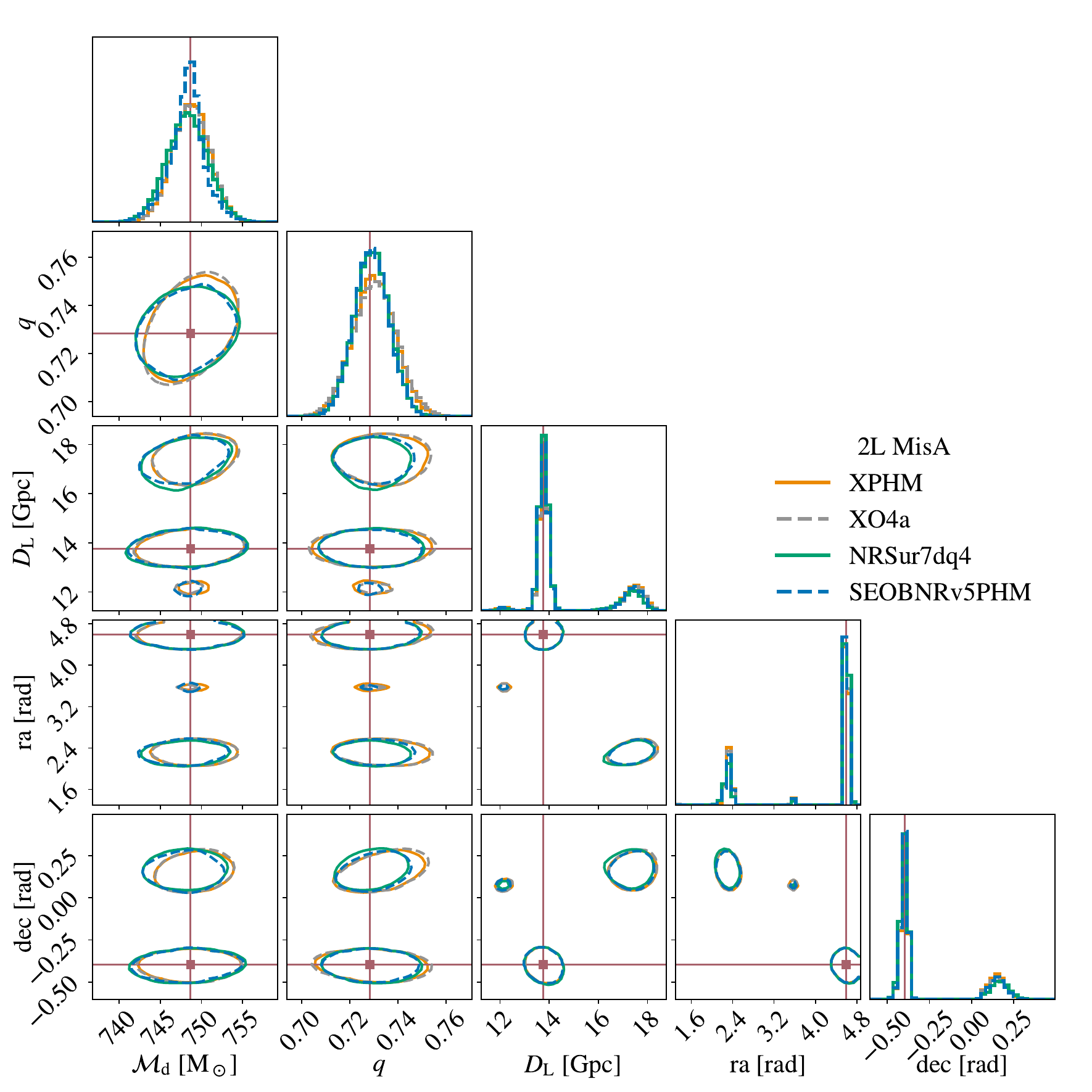}
    \caption{Marginalized one- and two-dimensional posterior distributions for the detector-frame chirp mass ($\mathcal{M}_\mathrm{d}$), mass ratio ($q$), luminosity distance ($D_\mathrm{L}$), right ascension (ra), and declination (dec) of the event displayed in Fig.~\ref{fig:dingo_bilby_2L_short}, observed with the 2L~MisA configuration in zero noise. Results from {\sc{Bilby}} are presented for different waveform approximants: {\sc{XPHM}} (orange), {\sc{XO4a}} (grey), {\sc{NRSur7dq4}} (teal), and {\sc{SEOBNRv5PHM}} (blue). Vertical and horizontal lines indicate the true injected values (see Table~\ref{tab:jsd}) and the contours denote the 95\% credible regions. Further details are provided in Appendix~\ref{app:waveforms}.}
    \label{fig:waveform}
\end{figure*}

To assess whether the multimodalities in luminosity distance and sky localization observed in Fig.~\ref{fig:dingo_bilby_2L_short} for the 2L~MisA configuration persists under different waveform approximants, we repeat the analysis in zero noise using {\sc{Bilby}} (see Appendix~\ref{app:bilby}). In addition to the fiducial {\sc{IMRPhenomXPHM}} waveform model ({\sc{XPHM}}; \citealt{Colleoni:2024knd};  see also Appendix~\ref{app:prior}), we both injected and recovered the signals using three additional state-of-the-art inspiral--merger--ringdown (IMR) approximants: {\sc{NRSur7dq4}} \citep{Varma:2019csw}, {\sc{SEOBNRv5PHM}} \citep{Ramos-Buades:2023yhy}, and {\sc{IMRPhenomXO4a}} \citep[{\sc{XO4a}}]{Thompson:2023ase}.

All considered waveform models describe precessing quasi-circular binaries and include higher-order multipole contributions; however, they adopt distinct modeling strategies \citep{Chatziioannou:2024hju}. In brief, {\sc{NRSur7dq4}} is constructed by interpolating numerical-relativity simulations \citep{Field:2013cfa,Blackman:2015pia} and is therefore typically the most accurate for high-mass systems, such as GW231123 \citep{LIGOScientific:2025rsn,Bini:2026kwz}. In contrast, {\sc{XPHM}}, {\sc{SEOBNRv5PHM}}, and {\sc{XO4a}} combine analytical frameworks with numerical-relativity calibration to build complete IMR models that are applicable across the full range of total masses \citep{Buonanno:1998gg,Buonanno:2000ef,Buonanno:2007pf,Ajith:2009bn}. The differing modeling strategies and treatments of precession dynamics render these waveform models largely independent. As shown in Fig.~\ref{fig:waveform}, we find no evidence of waveform-induced systematics in the inferred parameters, with all four waveform models reproducing the same number and locations of multimodalities.

\section{Sky modes in the 2L misaligned configuration}
\label{sec:time}

\begin{figure*}
    \centering
    \includegraphics[width=0.9\linewidth]{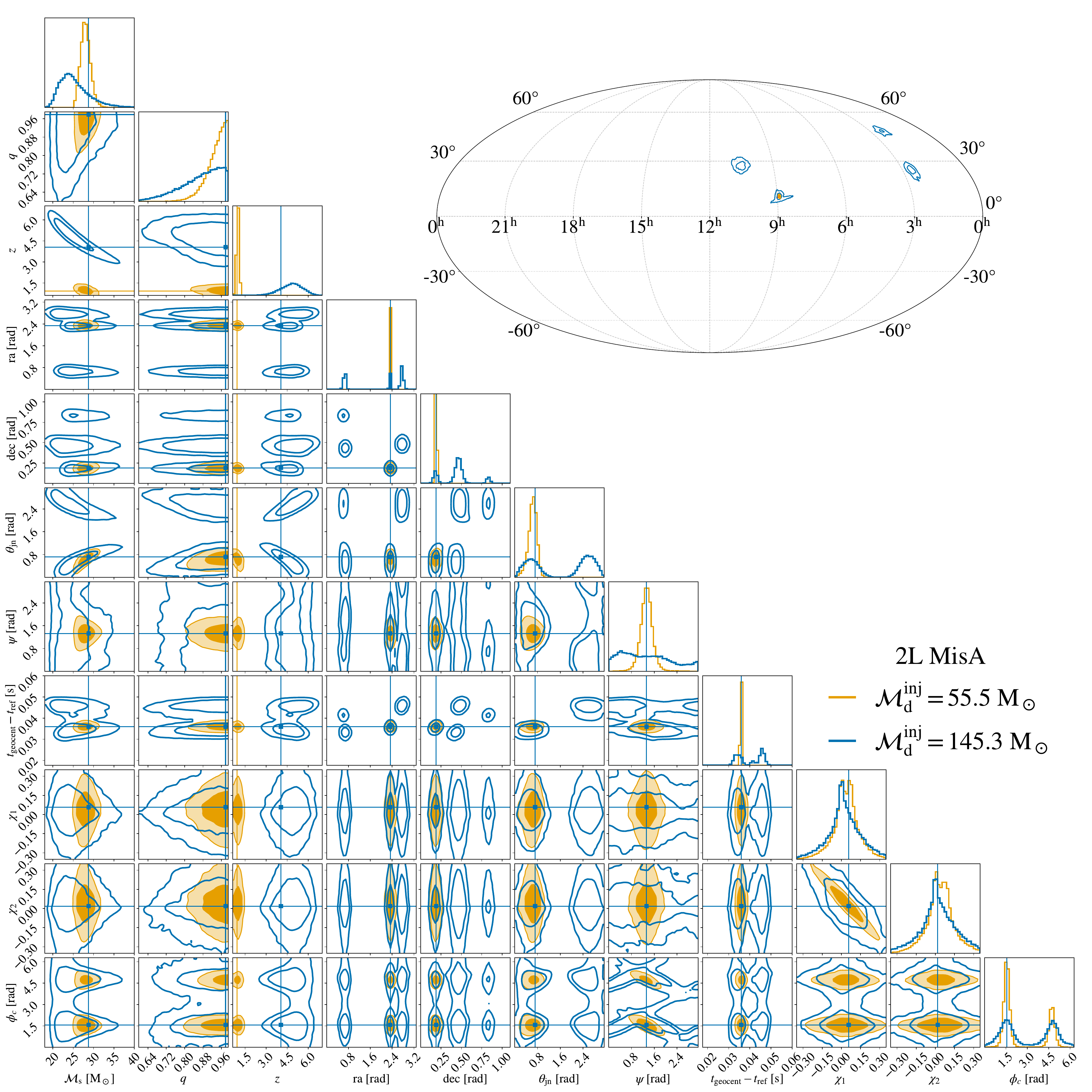}
    \caption{Marginalized one- and two-dimensional posterior distributions for all parameters, with the exception that this time we show the source-frame chirp mass ($\mathcal{M}_\mathrm{s}$) and redshift ($z$). The results for the low-mass event (orange) are shown alongside those for the high-mass event (blue). Vertical and horizontal lines indicate the true injected values and contours represent the 68\% and 95\% credible regions. \textit{Top-right panel:} Sky map of the low-mass (orange) and high-mass (blue) events, with contours indicating the 50\% and 90\% credible area. See Appendix~\ref{sec:time} for further details.}
    \label{fig:HM_LM}
\end{figure*}

Figure~\ref{fig:dingo_bilby_2L_short} illustrates multimodalities in sky localization 
for a source observed with the 2L~MisA configuration. The presence of these multimodalities is further confirmed in Fig.~\ref{fig:modes}, where almost no sources exhibit a single mode. Most sources ($\gtrsim 60\%$) display between two and six sky modes, while $\sim 20\%$ exhibit eight distinct modes. 

In this section we provide an intuitive example to illustrate that the presence of multiple sky modes in the 2L~MisA configuration is specific to the sources considered in this study, after applying the sample efficiency selection (Sect.~\ref{sec:high_eff} and the bottom panel of Fig.~\ref{fig:SNR_vs_eff}). For a small subset of sources that are excluded from our analysis but still present in the prior space, it is reasonable to expect that the number of sky modes would be significantly lower. Intuitively, the baseline between the 2L ET interferometers becomes effectively long for sources with sufficiently high S/N above a certain frequency threshold, which depends on both the length of baseline and the source’s sky location. A comprehensive study to quantify this effect is beyond the scope of this manuscript and will be addressed in future work.

Figure~\ref{fig:HM_LM} shows the one- and two-dimensional marginal distributions obtained with {\sc Bilby} for two sources at zero noise: a less massive, closer source with $\mathcal{M}^{\mathrm{inj}}_\mathrm{d} = 55.5~\mathrm{M}_\odot$ and $D^{\mathrm{inj}}_\mathrm{L} = 6.2~\mathrm{Gpc}$, with optimal S/N = 87.5, and a more massive, more distant source with $\mathcal{M}^{\mathrm{inj}}_\mathrm{d} = 145.3~\mathrm{M}_\odot$ and $D^{\mathrm{inj}}_\mathrm{L} = 37.2~\mathrm{Gpc}$ with optimal S/N = 28.7. The sky location, source-frame chirp mass, mass ratio, and other parameters are fixed for both sources to the maximum-likelihood values estimated for GW250114 \citep{LIGOScientific:2025rid}, while only the luminosity distance---and thus the redshift---are varied. Figure~\ref{fig:HM_LM} illustrates that multimodalities in sky localization vanish for the less massive source, whereas they persist for the more massive one, yielding $\Delta \Omega_{90\%} = 3$ deg$^2$ for the low-mass event and $\Delta \Omega_{90\%} = 194$ deg$^2$ for the high-mass event.



\end{appendix}
\end{document}